\documentclass{sn-jnl}

\jyear{2022}%

\usepackage{rotating}
\usepackage{graphicx}
\usepackage{dcolumn}
\usepackage{bm}
\usepackage{comment}
\usepackage{graphicx}
\usepackage{epsfig}
\usepackage{dcolumn}
\usepackage{caption}
\usepackage{mathrsfs}
\usepackage{float}
\usepackage{dsfont}	
\usepackage{relsize}
\usepackage{verbatim}
\usepackage{hyperref}
\usepackage{epstopdf}
\usepackage{amsfonts}
\newcommand{\Hsquare}{%
  \text{\fboxsep=-.2pt\fbox{\rule{0pt}{1ex}\rule{1ex}{0pt}}}%
}
\usepackage{float}
\usepackage{ragged2e}
\usepackage{setspace}
\usepackage{amsmath,amssymb,amsfonts}
\usepackage{amsmath}
\usepackage{amssymb}
\usepackage{physics}
\usepackage{dsfont}
\usepackage{mathrsfs}
\usepackage{wasysym}
\usepackage{graphicx}
\usepackage{textcomp}

\begin{document}

\title[Discrete-time Quantum Walks in Qudit Systems]{Discrete-time Quantum Walks in Qudit Systems}


\author*[1,2]{\fnm{Amit} \sur{Saha}}\email{abamitsaha@gmail.com}

\author[2]{\fnm{Debasri} \sur{Saha}}\email{debasri\_cu@yahoo.in}
\equalcont{These authors contributed equally to this work.}

\author[2]{\fnm{Amlan} \sur{Chakrabarti}}\email{acakcs@caluniv.ac.in}
\equalcont{These authors contributed equally to this work.}

\affil*[1]{\orgname{Atos}, \orgaddress{ \city{Pune}, \postcode{411045}, \state{Maharashtra}, \country{India}}}

\affil[2]{\orgdiv{A. K. Choudhury School Of Information Technology}, \orgname{University of Calcutta}, \orgaddress{\city{Kolkata}, \postcode{700106}, \state{West Bengal}, \country{India}}}


\abstract{Quantum walks contribute significantly to developing quantum algorithms and quantum simulations. Here, we introduce a first of its kind one-dimensional quantum walk in the $d$-dimensional quantum domain, where $d>2$, and show its equivalence for circuit realization in an arbitrary finite-dimensional quantum logic for utilizing the advantage of larger state space, which helps to reduce the run-time of the quantum walks as compared to the conventional binary quantum systems. We provide efficient quantum circuits for the implementation of discrete-time quantum walks (DTQW) in one-dimensional position space in any finite-dimensional quantum system when the dimension is odd using an appropriate logical mapping of the position space on which a walker evolves onto the multi-qudit states. With example circuits for various qudit state spaces, we also explore scalability in terms of $n$-qudit $d$-ary quantum systems. Further, the extension of one-dimensional DTQW to $d$-dimensional DTQW using $2d$-dimensional coin space on $d$-dimensional lattice has been studied, where {$d\ge2$}. Thereafter, the circuit design for the implementation of scalable $d$-dimensional DTQW in $d$-ary quantum systems has been portrayed. Lastly, we exhibit the circuit design for the implementation of DTQW using different coins on various search spaces.}




\maketitle

\section{Introduction}

 Recently, the development of quantum computers has achieved phenomenal progress, so researchers have shown a remarkable interest in implementing quantum algorithms, which are asymptotically superior as compared to its equivalent classical algorithms \cite{nielsen_chuang_2010, Farhi_1998}. Quantum walk \cite{PhysRevA.48.1687}, the counterpart of classical random walks, is one such quantum algorithm that has been an excellent candidate to constitute a universal model of quantum computation. Quantum Walk (QW) has also exhibited as a quantum search algorithm for the last two decades for its numerous speedups and applications on computational search problems \cite{Shenvi_2003, ambainis2004quantum, aharonov2000quantum, magniez2003quantum}. Quantum walks have two main variants, one is discrete-time quantum walks (DTQW) \cite{ambainis2003quantum, Tregenna_2003} and another is continuous-time quantum walks (CTQW) \cite{Childs_2003}. In this article, we are mainly focused on the discrete variant of quantum walks. The discrete-time quantum walks are usually defined on the combination of the coin (particle) and position Hilbert space, which spreads quadratically faster in position space in comparison to classical random walks \cite{Kempe03, Venegas_Andraca_2012, Kendon_2006}. The evolution of this position space is driven by a position shift operator controlled by a coin flip operator, such as a two-state Hadamard coin flip operator on line. 

In this article, efficient quantum circuits for the implementation of DTQW on varied search spaces in qudit systems have been proposed. Although physical systems in classical hardware are typically binary, common quantum hardware, such as in superconducting \cite{32} and trapped ion computers \cite{31, Ringbauer_2022}, has an infinite spectrum of discrete energy levels. Quantum hardware may be configured to manipulate the $d$ energy levels by operating on $d$-ary quantum systems. Qudit gates have already been successfully implemented \cite{1, 6, 13, 17, 19, 34} indicating it is possible to consider higher-level systems apart from qubit-only systems. Thus, the question of higher states, beyond the standard two being implemented and performed, no longer stands strong \cite{sahapra}. Therefore, we graduate to multi-valued quantum systems or qudits \cite{Muthukrishnan_2000}, which in the course reduce the circuit complexity and commend the efficiency of quantum algorithms \cite{9410395} to provide larger state space with simultaneous multiple control operations \cite{Wang_2020, qft, Bocharov_2017}. For instance, $N$ qubits can be formulated as $\frac{N}{\log_2 {d}}$ qudits, which immediately gives $\log_2 {d}$-factor in run-time \cite{Fan_2007, Khan_2006, Di_2013} for solving any computational problem using quantum algorithm.

  Many different quantum walks-related experiments have been carried out on real quantum hardware \cite{balu2017physical, Yan753, alderete2020quantum, acasiete2020experimental}. Albeit apart from the efficient implementation of one-dimensional DTQW in binary \cite{singh2020universal} and ternary quantum systems \cite{9410395}, efficient implementation of DTQW in an arbitrary finite-dimensional quantum system still remains arduous \cite{Zhou_2019}. For an arbitrary finite-dimensional quantum computer or hardware, the main limitations are the number of qubits/qudits and the coherence time of the system which limits the probable steps of DTQW that can be implemented. The challenge is to utilize the qubits/qudits in such a manner that the maximum possible number of steps of DTQW can be implemented with a minimum number of qubits/qudits. This encouraged us to carry out the efficient circuit realization of DTQW with the help of the nearest-neighbor position mapping approach in an arbitrary finite-dimensional quantum system in this article. 
  
  The circuit realization is such that whenever $d$-ary quantum computers are in use, we can map them to it right away to gain an edge in quantum walk applications on search problems \cite{santha, ambainis05}. We show that the {$\lceil\log_{d}(2n+1)\rceil$} qudits are enough for the implementation of $n$-step one-dimensional DTQW and can further be scaled up to implement more steps. To the best of our knowledge, there is no efficient generalized implementation of higher-dimensional discrete-time quantum walks in an arbitrary finite-dimensional quantum system, which has enormous search applications in the literature \cite{ambainis2011search, Rhodes_2020}. Hence, we also apply the nearest-neighbor approach for higher-dimensional discrete-time quantum walks so that we can implement higher-dimensional DTQW by scaling the dimensionality of the proposed efficient implementation of one-dimensional DTQW when the dimension is odd. Our circuit design can be implemented in any qudit-supported quantum hardware since all the qudit gates that are used in our design are universal, which makes our work generalized in nature.

In this article, we have implemented DTQW on different search-space using various coins in qudit systems. Our novelty lies in the fact that:

\begin{itemize}
    
    \item We realize quantum circuit to implement one-dimensional DTQW in $d$-dimensional quantum settings for the first of its kind.
     
    \item We also address the scalability of the proposed circuit in terms of $n$-qudit systems, which makes this circuit realization generalized in nature for implementing more steps. We show generalized implementation of DTQW based on nearest-neighbor approach for any finite-dimensional system when the dimension is odd and generalized implementation of DTQW based on increment-decrement approach for any finite-dimensional system when the dimension is even.
    
    \item We further define $d$-dimensional DTQW using $2d$-dimensional coin on $d$-dimensional lattice in an arbitrary finite-dimensional quantum system, where $d \ge 2$ for the first time to the best of our knowledge.
    
    \item We portray an efficient quantum circuit realization to implement $d$-dimensional DTQW in $d$-dimensional quantum settings with an example circuit for two-dimensional DTQW for the first of its kind using an appropriate logical mapping of the position space on which a walker evolves onto the multi-qudit states.
    
    \item Lastly, we efficiently simulate DTQW for different coin operators considering various position spaces.
\end{itemize}

{The structure of this article is as follows. Section \ref{2} describes the dynamics of one-dimensional and higher-dimensional DTQW. Section \ref{2} also describes the state-of-the-art implementation of one-dimensional DTQW in binary and ternary quantum systems. Section \ref{3} proposes efficient quantum circuit implementation for one-dimensional DTQW in $d$-dimensional quantum systems followed by the generalization of the quantum circuit. Section \ref{3} also proposes efficient quantum circuit implementation for $d$-dimensional quantum walks on $d$-dimensional lattice in $d$-dimensional quantum systems. Section \ref{4} analyzes the efficiency of the proposed position state mapping onto the multi-qudit states. Section \ref{5} captures our conclusions.}

\section{Background}\label{2}
In this section, firstly, we have discussed qudits and generalized quantum gates, which are used for implementing DTQW. Then, we move towards a brief discussion on one-dimensional DTQW, especially the state-of-the-art efficient implementation of one-dimensional DTQW in binary \cite{singh2020universal} and ternary quantum systems \cite{9410395}. These state-of-the-art efficient implementations using the nearest-neighbor approach laid the foundation of our proposed implementation of one-dimensional DTQW in an arbitrary finite-dimensional qudit system. Lastly, we put some light on higher-dimensional DTQW in a higher-dimensional lattice structure.

\subsection{Quantum circuit}
Any quantum algorithm can be expressed or visualized in the form of a quantum circuit. Commonly binary quantum systems, logical qubits, and quantum gates comprise these quantum circuits \cite{barenco}. The number of gates present in a circuit is called gate count and the number of qubits present in a circuit is known as qubit cost. In this work, we mainly deal with qudits and generalized quantum gates.
\subsubsection{Qudits}
 Logical qudit that encodes the input/output of a quantum algorithm in multi-valued quantum systems is often referred to as data qudit. Ancilla qudit is another type of qudit used to store temporary results. In $d$-dimensional quantum systems \textit{qudit} is the unit of quantum information. Qudit states can be manifested as a
vector in the~$d$ dimensional Hilbert space~$\mathscr{H}_d$.
 The span of orthonormal basis vectors $\{\ket{0},\ket{1},\ket{2},\dots \ket{d-1}\}$ is the vector space. 
In a qudit system, the general form of a quantum state  can be expressed as 
\begin{equation}
\ket{\psi}=\alpha_0 \ket0 +\alpha_1 \ket1 +\alpha_2 \ket2+\cdots+\alpha_{d-1} \ket{d-1}=
\begin{pmatrix}
\alpha_0 \\
\alpha_1 \\
\alpha_2 \\
\vdots   \\
\alpha_{d-1} \\
\end{pmatrix}
\end{equation}
where $\vert\alpha_0\vert^2+\vert\alpha_1\vert^2+\vert\alpha_2\vert^2+\cdots+\vert\alpha_{d-1}\vert^2=1$ and $\alpha_0$, $\alpha_1$, $\dots$, $\alpha_{d-1} \in\mathbb{C}^d$.

\subsubsection{Generalized quantum gates}

 In this section, an outline of generalized qudit gates \cite{Wang_2020, Garcia_Escartin_2013} is conferred. The generalization can be delineated as discrete quantum states of any arity.  In a quantum algorithm, for modification of the quantum state, unitary qudit gates are applied. For the logic synthesis of quantum walks in $d$-dimensional quantum system, it is necessary to take into account one-qudit generalized gates viz. NOT gate ($X_d$),  Hadamard  gate ($F_d$), two-qudit generalized CNOT gate ($C_{X,d}$) and Generalized multi-controlled Toffoli gate ($C^{n}_{X,d}$). These gates are expressed in detail for better understanding:

\textbf{Generalized NOT gate:} $X^d_{+a}$, the generalized NOT can be defined as $X^d_{+a}\ket{x}=\ket{(x+a) \mod d}$, where $1 \le a \le d-1$. For visualization of the $X^d_{+a}$ gate, we have used a 'rectangle' ($\Hsquare$). '$X^d_{+a}$' in the 'rectangle' box represents the generalized NOT. In binary quantum systems, NOT gate is represented as $\otimes$ or '$X$' in the '$\Hsquare$' box.

\textbf{Generalized Hadamard gate:} $F_d$, the generalized quantum Fourier transform or generalized Hadamard gate, produces the superposition of the input basis states. We have used $F_d$ in the 'rectangle' ($\Hsquare$) box to represent the generalized Hadamard gate. The $(d \times d)$ matrix representation of it is as shown below  
:

\begin{align*}
F_d = {1\over\sqrt{d}} \left(\begin{matrix} 1 & 1 & 1 & \ldots & 1 \\ 1 & \omega & \omega^2 & \ldots & \omega^{d-1}  \\ 1 & \omega^2 & \omega^4 & \ldots & \omega^{2(d-1)}  \\ \vdots & \vdots & \vdots & \ddots & \vdots \\ 1 & \omega^{d-1} & \omega^{2(d-1)} & \ldots & \omega^{(d-1)(d-1)} \end{matrix}\right)
  \end{align*} 

In binary quantum systems, we have used $H$ in the 'rectangle' ($\Hsquare$) box to represent the Hadamard gate.

\textbf{Generalized CNOT gate:} Quantum entanglement is an unparalleled property of quantum mechanics and can be attained by a controlled-NOT (CNOT) gate in a binary quantum system. For $d$-dimensional quantum systems, the binary 2-qubit CNOT gate is generalized to the $INCREMENT$ gate:\\ $\text{INCREMENT}\ket{x}\ket{y}=\ket{x}\ket{(y+a) \mod d}$, if $x=d-1$, and = $\ket{x}\ket{y}$, otherwise, where $1 \le a \le d-1$. In the schematic design of the generalized CNOT gate, $C_{X,d}$, we have used a 'Black dot' ($\bullet$) to represent the control, and a 'rectangle' ($\Hsquare$) to represent the target. '$X^d_{+a}$' in the target box represents the increment operator. 

In binary quantum systems, we have used a 'Black dot' ($\bullet$) to represent the control, and $\oplus$ to represent the target for representing the CNOT gate.

\textbf{Generalized Multi-controlled Toffoli gate:} We extend the generalized CNOT or $INCREMENT$ further to operate over $n$ qudits as a generalized Multi-controlled Toffoli Gate or $n$-qudit Toffoli gate $C_{X,d}^n$.  For $C_{X,d}^n$, the target qudit is increased by $a \ (\text{mod } d)$ only when all $n-1$ control qudits have the value $d-1$, where $1 \le a \le d-1$. In the schematic design of the generalized Multi-controlled Toffoli Gate, $C_{X,d}^n$, we have used a 'Black dot' ($\bullet$) to represent all the control qudits, and a 'rectangle' ($\Hsquare$) to represent the target. '$X^d_{+a}$' in the target box represents the increment operator. In binary quantum systems, we have used a 'Black dot' ($\bullet$) to represent all the control qubits, and $\oplus$ to represent the target for representing the Multi-controlled Toffoli (MCT) gate. 

Owing to technology constraints, a multi-controlled Toffoli gate can be substituted by an equivalent circuit comprising one-qudit and/or two-qudit gates. As discussed in \cite{sahapra}, irrespective of the quantum systems, these multi-controlled gates can be decomposed with the same gate-cost and circuit-depth without ancilla in any finite-dimensional quantum system with the help of intermediate qudits. In fact, this decomposition only requires generalized CNOT gates, no T gate or S gate is required. Hence, to implement DTQW, qudit systems will be beneficial over qubit systems concerning qudit-cost, gate-cost, and circuit-depth, since, while using qudit systems over binary systems, the less number of qudits are required to implement the same number of steps of DTQW as compared to the number of qubits.

\subsection{One-dimensional discrete-time quantum walks}

Discrete-time quantum walks take place in the product space $\mathcal{H}={\mathcal{H}}_{p}\bigotimes {\mathcal{H}}_{c}$. ${\mathcal{H}}_{p}$ is a Hilbert space which has an orthonormal basis given by the position states $\{\vert x\rangle, x\in\mathcal{Z}\}$. The default initial position state is $\vert0\rangle$. Due to the two choices of movement, one-dimensional DTQW has a two-dimensional coin. Therefore, $\mathcal{H}_{c}$ is a Hilbert space spanned by the orthonormal basis$\{\ket{\uparrow},  \ket{\downarrow}\}$ ($\uparrow$ for right and $\downarrow$ for left).

Let $\vert x, \alpha\rangle$ be a basic state, where $x\in \mathcal{Z}$ represents the position of the particle in binary quantum systems and $\alpha\in \{\uparrow,  \downarrow\}$ represents the coin state, more specifically, in this paper, $\uparrow$ corresponds to $\ket{0}$ and $\downarrow$ corresponds to $\ket{1}$. The evolution of the whole system at each step of the walk can be described by the unitary operator denoted by $U$,
\begin{equation}\label{21}
    U=S(I \otimes C),
\end{equation} where $S$ is the shift operator defined by
\begin{equation} \label{Shift}
S  = \sum_{x\in\mathbb{Z}} \bigg (\ket{\uparrow}\bra{\uparrow}
\otimes   \ket{x+1}\bra{x}+\ket{\downarrow}\bra{\downarrow} \otimes \ket{x-1}\bra{x}\bigg ).
\end{equation}

$I$ is the identity matrix which operates in $\mathcal{H}_{p}$, while $C$ is the coin operation. Hadamard ($H$) coin is an example of a two-dimensional coin flip operator, which is denoted by $C$ here,
\begin{equation}
    C=\frac{1}{\sqrt{2}}\left(
          \begin{array}{ccc}
            1 & 1 \\
            1 & -1 \\
          \end{array}
        \right).
\end{equation}

Similarly, one-dimensional DTQW in qudit systems can be carried out by spanning the one-dimensional position Hilbert space in higher dimensional qudit systems to represent the labels on the position states in qudit systems. Likewise, in this article, we try to efficiently implement one-dimensional DTQW in $d$-dimensional quantum systems. We also exhibit quantum circuits for implementing the one-dimensional quantum walks in these quantum systems more elaborately in Appendix \ref{appen1} and \ref{appen2}.

\subsection{Higher-dimensional discrete-time quantum walks}

As discussed, discrete-time quantum random walks consist of a position Hilbert space $H_p$ and a coin Hilbert space $H_c$. A quantum state consists of these two degrees of freedom, $\ket{c} \otimes \ket{v}$ where $\ket{c} \in H_c$ and $\ket{v} \in H_p$. A step in quantum walks is a unitary evolution $U = S.(C \otimes I)$ where $S$ is the shift operator and $C$ is the coin operator, which acts only on the coin Hilbert space $H_c$. For higher-dimensional, such as two-dimensional DTQW, if we consider a $\sqrt{N} \times \sqrt{N}$ grid, then the quantum walk starts in a superposition of states given by
\begin{equation}
    \ket{\psi(0)} = \frac{1}{\sqrt{4N}}(\sum_{i=1}^{4}\ket{i} \otimes \sum_{x,y=1}^{\sqrt{N}}\ket{x,y})
\end{equation}

where (i) each location $(x,y)$ corresponds to a quantum register $\ket{x,y}$ with $x,y \in \{1, 2, \hdots, \sqrt{N}\}$ and (ii) the coin register $\ket{i}$ with $i \in \{\leftarrow, \rightarrow, \uparrow, \downarrow\}$. The most often used transformation on the coin register is Grover's Diffusion transformation $D$

\begin{equation}
\label{eq:diffusion}
    D = \frac{1}{2}\begin{pmatrix}
    -1 & 1 & 1 & 1\\
    1 & -1 & 1 & 1\\
    1 & 1 & -1 & 1\\
    1 & 1 & 1 & -1
    \end{pmatrix}
\end{equation}

The Diffusion operator can also be written as $D = 2\ket{s_D}\bra{s_D} - I_4$, where $\ket{s_D} = \frac{1}{\sqrt{4}}\sum_{i=1}^{4}\ket{i}$.\\

The transformation creates a superposition of the coin states $\ket{i}$, which in turn governs the shift operation. Multiple shift operators have been proposed in the literature, out of which, in this article, we have used the  shift transformation $S$ \cite{ambainis2004coins} whose action on the basis states are as follows:
\begin{align}
    \ket{i,j,\uparrow} & = \ket{i,j-1,\uparrow}\\
    \ket{i,j,\downarrow} & = \ket{i,j+1,\downarrow}\\
    \ket{i,j,\leftarrow} & = \ket{i-1,j,\leftarrow}\\
    \ket{i,j,\rightarrow} & = \ket{i+1,j,\rightarrow}.
 \end{align}

 These position states can be implemented in an arbitrary finite-dimensional quantum system by spanning the position Hilbert space $H_p$ in an arbitrary finite-dimensional quantum system. The periodicity in two-dimensional DTQW has important applications in quantum algorithms \cite{ambainis2011search}, especially in the search algorithms by designing the structure of the grid and the initial state of the walker in a special manner, which is not a subject matter of this paper, hence, it remains a future scope of this paper.

Like two-dimensional DTQW, higher-dimensional DTQW or $d$-dimensional DTQW can be carried out in $d$-dimensional lattice by increasing the dimension of the coin operator to $d$. This $d$-dimensional DTQW can also be efficiently implemented in an arbitrary finite-dimensional quantum system by spanning the position Hilbert space $H_p$ in an arbitrary finite-dimensional quantum system, which is shown in this article.

\section{ Quantum circuit for implementing the one-dimensional quantum walks in qudit systems}\label{3}

This section exhibits the proposed quantum circuit for implementing the one-dimensional DTQW in qudit systems. From now on, for simplicity's sake, we consider qubit systems to represent the coin states. It is to be noted that the construction of the proposed quantum circuit for one-dimensional DTQW in qudit systems is similar to all odd-dimensional qudit systems and is also akin to all even-dimensional qudit systems.  

\subsection{ Quantum circuit for implementing the one-dimensional quantum walks in $d$-ary quantum systems when $d$ is odd}

 In this subsection, we present the logical realization of quantum circuits to implement one-dimensional quantum walks using two state coins in $d$-ary quantum systems, when $d$ is odd. As an example, we have taken 5-ary and 7-ary quantum systems. For 5-ary quantum systems, to implement a DTQW in a one-dimensional position Hilbert space of size $5^q$, $(q)$ qudits of 5-ary quantum systems and one qubit
are required, one qubit to represent the particle's internal state ({coin qubit}) and $q$-qudits to represent its position. As discussed, the coin operation can be implemented by applying a single qubit rotation gate on the coin qubit, and the position shift operation is implemented subsequently with the help of multi-qudit gates where the coin qubit acts as the control. Quantum circuits for implementing DTQW depend on how the position space is represented. Example circuits for a four-qubit-qudit system are given in this section.

For $q=3$ the number of steps of DTQW that can be implemented is $\lfloor5^{q}/2\rfloor = 62$. We choose the position state mapping given in Table \ref{tab_5_dtqw}, \ref{tab_5_dtqw1}, \ref{tab_5_dtqw2} with a fixed initial position state $\ket{000}$. Fixing the initial state of the walker helps in reducing the gate count in the quantum circuit and hence reduces the overall error. For example, if the initial state is not fixed to $\ket{000}$ then at first we have to bring the initial state to $\ket{000}$ with the help of 1-qudit generalized NOT gates. We denote the initial state as $\ket{x = 0} \equiv \ket{000}$ in Table \ref{tab_5_dtqw}. After each step of the DTQW, two new position states must be considered. In Table \ref{tab_5_dtqw}, \ref{tab_5_dtqw1}, \ref{tab_5_dtqw2}, we show that the mapping of these new position states onto the multi-qudit states is in such a way that efficient number of gates are used to implement the shift operation, we consider the nearest-neighbor position space so as to make the circuit efficient.

After the first step of DTQW, if the coin state is $\ket{0}$, the particle moves to the right, $\ket{x = 1} \equiv \ket{001}$, if the coin state is $\ket{1}$, the particle moves to the left, $\ket{x = -1} \equiv \ket{004}$ as shown in Table \ref{tab_5_dtqw}. We are following the nearest-neighbor approach to determine the position states as shown in Table \ref{tab_5_dtqw} so that the least significant qudit (LSQ) only changes, rest of the two qudits remain unchanged. The mathematical formulation of this step can be described as follows:

Suppose at $t=0$, initial state: $\ket{0} \otimes \ket{000}$ 

At $t=1$, after first step:  $a_0 * \ket{0} \otimes \ket{001} + b_0 * \ket{1} \otimes \ket{004}$, where $a_0$ and $b_0$ are the amplitudes.

\begin{figure}[ht!]
\centering   
\includegraphics[width=6cm, scale=1]{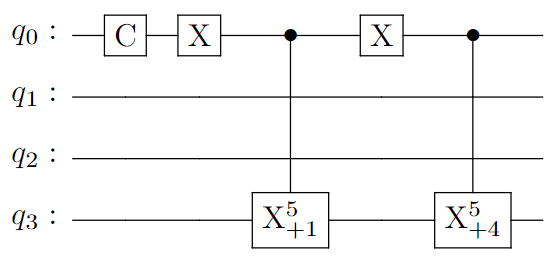}
    \caption{Quantum circuit for the first two steps of DTQW on three qudits in 5-ary quantum systems as given in Table \ref{tab_5_dtqw}.}
    \label{fig_5_dtqw}
\end{figure}

This logic is mapped into the circuit as shown in Figure \ref{fig_5_dtqw}. In Figure \ref{fig_5_dtqw}, the first qubit is used for coin operation and the rest of the three qudits are for position states on which shift operation will be performed as per the coin state. Firstly, all four qubit-qudits are initialized with $\ket{0}$.

$\psi_0 {=} \ket{0} \otimes \ket{000}$

A two-state coin $C$ has been applied on the first qubit so that we can have a superposition state,

$\psi_1 {=} a_0 * \ket{0} \otimes \ket{000} + b_0 * \ket{1} \otimes \ket{000}$. 

Next, we have to perform the shift operation. For that, we have applied NOT gate on qubit so that we can perform shift operation for the coin state $\ket{0}$. Hence, the quantum state becomes,

$\psi_2 {=}$  $a_0 * \ket{1} \otimes \ket{000} + b_0 * \ket{0} \otimes \ket{000}$.  

For the shift operation for the coin state $\ket{0}$, we have applied a 2-qubit-qudit controlled-$X_{+1}^5$ gate on qubit as control and last qudit as a target. Now the quantum state evolves as

$\psi_3 {=}$  $a_0 * \ket{1} \otimes \ket{001} + b_0 * \ket{0} \otimes \ket{000}$.

To take back the coin state to its previous state for further operation, we have to apply the NOT gate on the qubit, such that the quantum state evolves,

$\psi_4 {=}$  $a_0 * \ket{0} \otimes \ket{001} + b_0 * \ket{1} \otimes \ket{000}$. 

Then, we applied a 2-qubit-qudit controlled-$X_{+4}^5$ gate on the qubit as control and the last qudit as the target for the shift operation for the coin state $\ket{1}$. Thus, the quantum state evolves,

$\psi_5 {=}$  $a_0 * \ket{0} \otimes \ket{001} + b_0 * \ket{1} \otimes \ket{004}$.

Similarly, we apply the same circuit for the next step of DTQW, so the quantum state evolves as,

$\psi_6 {=}$  $a_1 * \ket{0} \otimes \ket{002} + b_1 * \ket{1} \otimes \ket{000}$ +  $c_1 * \ket{0} \otimes \ket{000} + d_1 * \ket{1} \otimes \ket{003}$.

\begin{table}[ht!]
\centering
\caption{Position state mapping with the multi-qudits states for quantum circuits presented in Figure \ref{fig_5_dtqw}.}
\begin{tabular}{|m{11em} | m{0.005em}| m{11em}  | }
\hline
~~~$\ket{x = 0} \equiv \ket{000}$  &&   \\
\hline
~~~$\ket{x = 1} \equiv \ket{001}$ && ~~~$\ket{x = -1} \equiv \ket{004}$\\
\hline 
~~~$\ket{x = 2} \equiv \ket{002}$ && ~~~$\ket{x = -2} \equiv \ket{003}$\\
\hline 
\end{tabular}
\label{tab_5_dtqw}
\end{table}

\begin{figure}[ht!]
  \centering
\includegraphics[width=9cm, scale=1]{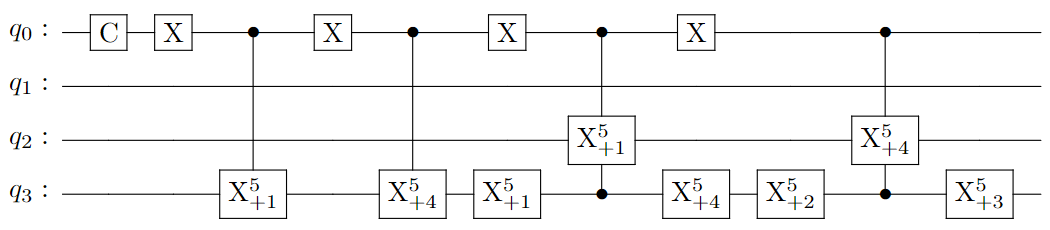}
    \caption{Quantum circuit for the third-twelfth steps of DTQW on three qudits in 5-ary quantum systems as shown in Table \ref{tab_5_dtqw1}.}
    \label{fig_5_dtqw1}
\end{figure}

As shown in the first two steps of DTQW, the same circuit works as only one qudit is required to represent the position states, since we are following the nearest-neighbor approach. We have observed that in the first step of DTQW only one qudit (last qudit) change was enough to describe the position states as after the first step of DTQW, there are only two different position states in the one-dimensional line. Similarly, for the next step, the change is again required only on the last qudit. For further third-twelfth DTQW steps, two qudits (third and last qudit) are sufficient to get all possible states to describe all the steps of DTQW as shown in Table \ref{tab_5_dtqw1}. Figure \ref{fig_5_dtqw1} describes the design of the quantum circuit for the third-twelfth DTQW steps as shown in Table \ref{tab_5_dtqw1}.

\begin{table}[ht!]
\centering
\caption{An example of position state mapping onto the multi-qudits states in 5-ary quantum systems for quantum circuits presented in Figure \ref{fig_5_dtqw1}.}
\begin{tabular}{|m{11em} | m{0.005em}| m{11em}  | }
\hline
~~~$\ket{x = 3} \equiv \ket{013}$ && ~~~$\ket{x = -3} \equiv \ket{042}$\\
\hline
~~~$\ket{x = 4} \equiv \ket{014}$  && ~~~$\ket{x = -4} \equiv \ket{041}$\\
\hline
~~~$\ket{x = 5} \equiv \ket{010}$ &&  ~~~$\ket{x = -5} \equiv \ket{040}$\\
\hline 
~~~$\ket{x = 6} \equiv \ket{011}$ && ~~~$\ket{x = -6} \equiv \ket{044}$ \\
\hline
~~~$\ket{x = 7} \equiv \ket{012}$ && ~~~$\ket{x = -7} \equiv \ket{043}$ \\
\hline
~~~$\ket{x = 8} \equiv \ket{023}$ && ~~~$\ket{x = -8} \equiv \ket{032}$ \\
\hline
~~~$\ket{x = 9} \equiv \ket{024}$ && ~~~$\ket{x = -9} \equiv \ket{031}$ \\
\hline
~~~$\ket{x = 10} \equiv \ket{020}$ && ~~~$\ket{x = -10} \equiv \ket{030}$ \\
\hline
~~~$\ket{x = 11} \equiv \ket{021}$ && ~~~$\ket{x = -11} \equiv \ket{034}$ \\
\hline
~~~$\ket{x = 12} \equiv \ket{022}$ && ~~~$\ket{x = -12} \equiv \ket{033}$ \\
\hline
\end{tabular}
\label{tab_5_dtqw1}
\end{table}

\begin{figure*}[ht!]
\centering   
\includegraphics[scale=0.5]{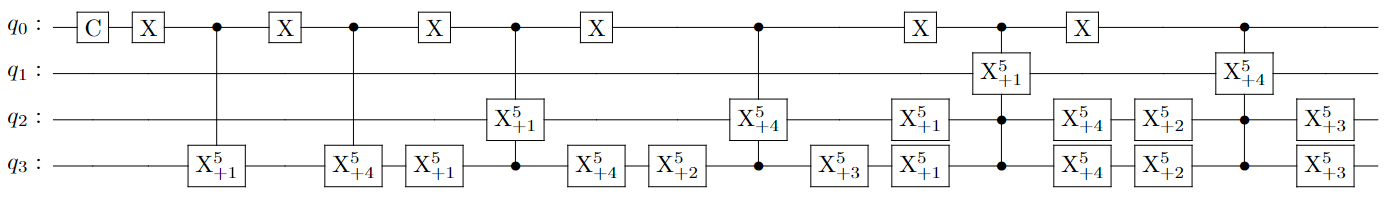}
    \caption{Quantum circuit for the thirteenth-twenty fourth steps of DTQW on three qudits in 5-ary quantum systems as shown in Table \ref{tab_5_dtqw2}.}
    \label{fig_5_dtqw2}
\end{figure*}

Now, for the thirteenth step of DTQW, two new position states are introduced. For that, the change in three qudits is required. At this stage to move forward, the circuit design for previous position states remains the same as shown in Figure \ref{fig_5_dtqw1}. Now, for further shift operation, we have to extend the circuit for $\ket{022}$ and $\ket{033}$ states as shown in Figure \ref{fig_5_dtqw2} to get a new position states $\ket{133}$ and $\ket{422}$ as shown in Table \ref{tab_5_dtqw2}. For further next ten steps of DTQW, this same circuit works as no new qudit is required to represent the position states since we are following the nearest-neighbor approach as shown in Table \ref{tab_5_dtqw2}.

\begin{table}[ht!]
\centering
\caption{An example of position state mapping onto the multi-qudits states in 5-ary quantum systems for quantum circuits presented in Figure \ref{fig_5_dtqw2}.}
\begin{tabular}{|m{11em} | m{0.005em}| m{11em}  | }
\hline
~~~$\ket{x = 13} \equiv \ket{133}$ && ~~~$\ket{x = -13} \equiv \ket{422}$ \\

\hline
~~~$\ket{x = 14} \equiv \ket{134}$ && ~~~$\ket{x = -14} \equiv \ket{421}$\\
\hline 
~~~$\ket{x = 15} \equiv \ket{130}$  &&  ~~~$\ket{x = -15} \equiv \ket{420}$ \\
\hline
~~~$\ket{x = 16} \equiv \ket{131}$ && ~~~$\ket{x = -16} \equiv \ket{424}$\\
\hline
~~~$\ket{x = 17} \equiv \ket{132}$  && ~~~$\ket{x = -17} \equiv \ket{423}$\\
\hline
~~~$\ket{x = 18} \equiv \ket{143}$ &&  ~~~$\ket{x = -18} \equiv \ket{412}$\\
\hline 
~~~$\ket{x = 19} \equiv \ket{144}$ && ~~~$\ket{x = -19} \equiv \ket{411}$ \\
\hline
~~~$\ket{x = 20} \equiv \ket{140}$ && ~~~$\ket{x = -20} \equiv \ket{410}$ \\
\hline
~~~$\ket{x = 21} \equiv \ket{141}$ && ~~~$\ket{x = -21} \equiv \ket{414}$ \\
\hline
~~~$\ket{x = 22} \equiv \ket{142}$ && ~~~$\ket{x = -22} \equiv \ket{413}$ \\
\hline
~~~$\ket{x = 23} \equiv \ket{103}$ && ~~~$\ket{x = -23} \equiv \ket{402}$ \\
\hline
~~~$\ket{x = 24} \equiv \ket{104}$ && ~~~$\ket{x = -24} \equiv \ket{401}$ \\
\hline
\end{tabular}
\label{tab_5_dtqw2}
\end{table}

\begin{figure}[ht!]
\centering
\includegraphics[width=8cm, scale=1]{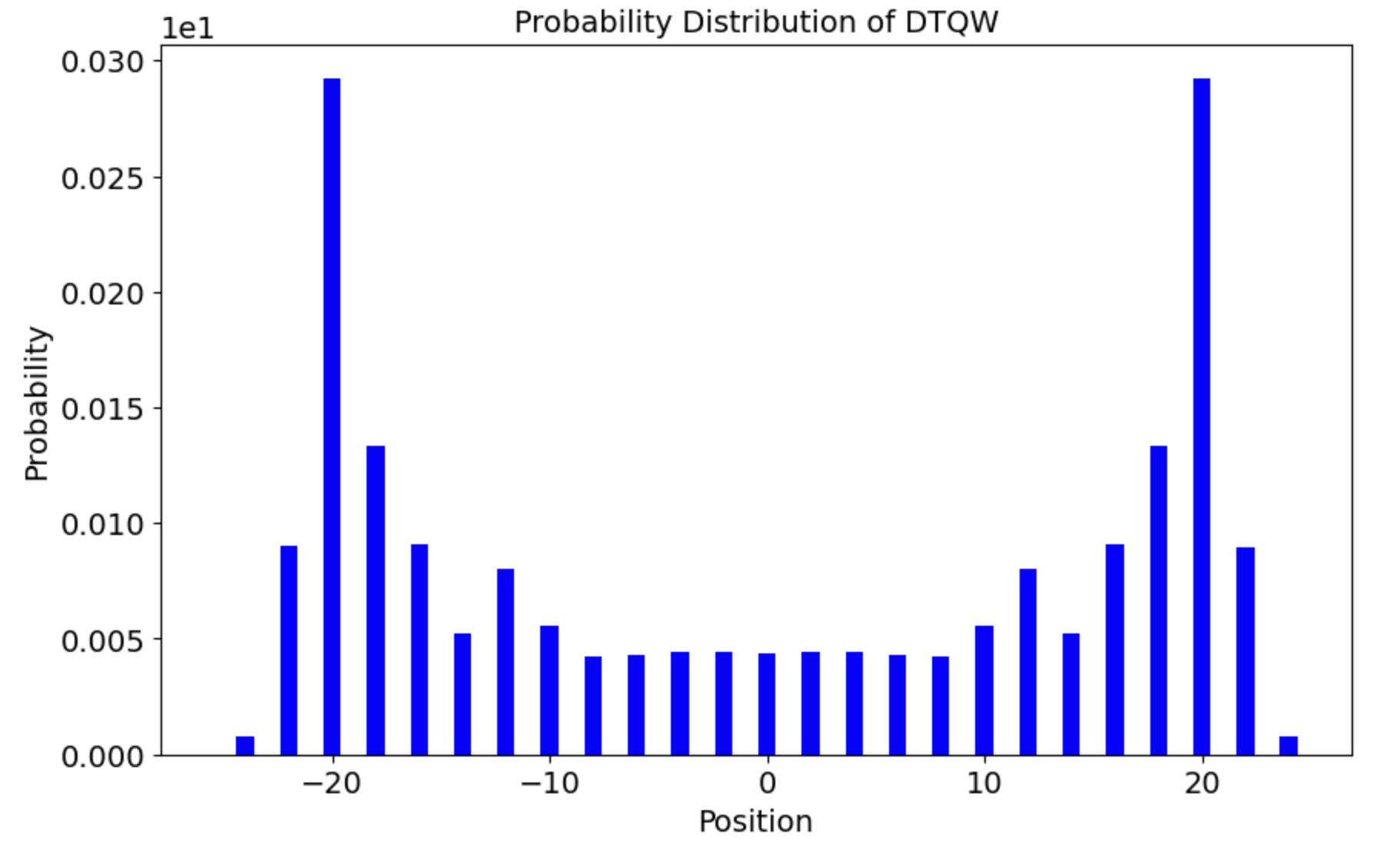}
\captionof{figure}{{Quantum walks on a one-dimensional line using a Hadamard coin, after 30 steps in 5-ary and 7-ary quantum systems. X-axis denotes the position states and the Y-axis denotes the probability.}}
\label{H_dtqw_5ary}
\end{figure}

{For a better understanding of position state mapping onto the multi-qudits states, we have considered another quantum system $i.e.,$ 7-ary quantum systems, which is illustrated in Appendix \ref{7-ary}.} Through numerical analysis, we have shown the probability distribution after 30 steps of a DTQW using a Hadamard coin for both the 5-ary and 7-ary quantum systems in Figure \ref{H_dtqw_5ary}.

\paragraph{Generalization of the quantum circuit for implementing one-dimensional discrete-time quantum walks in $d$-ary quantum systems when $d$ is odd}

{The proposed circuits can be scaled to implement more steps of one-dimensional DTQW on larger $d$-ary systems when $d$ is odd with the help of higher controlled generalized MCT gates. As shown in Table \ref{tab_5_gen}, using $n+1$-qubit-qudit systems, implementation of $\lfloor d^{n}/2\rfloor$-steps of a DTQW can be performed. In other words, to implement the $n$-step of DTQW,  $\lceil\log_{d}(2n+1)\rceil$ qudits $+ 1$ qubit are required. Through mathematical induction, it can be easily shown that for $n$-step of DTQW,  $(\lceil\log_d{(2n+1)}\rceil)+1$ qubit-qudits are required, where $\lceil\log_d{(2n+1)}\rceil$ qudits for position states and $1$ qubit for coin.}

\begin{table} [ht!]
\centering
\caption{Number of steps and a maximum number of control qudits and qubit required to control a target qudit in the DTQW for 5-ary systems of up to $n$ qudits.}
\scalebox{.8}{
\begin{tabular}{|m{6em}| m{6em} | m{6em} | m{7.5em} |}
\hline
No. of qudits & No. of qubit & No. of steps & Max. No. of controls in Generalized MCT gates\\
\hline
 1 & 1 & 2 & 1 \\
\hline
 2 & 1 & 12 & 2 \\
\hline 
 3 & 1 & 62 & 3\\
\hline
 4 & 1 & 312 & 4 \\
\hline
$n$ & 1 & $\lfloor 5^n/2 \rfloor$ & n\\
\hline
\end{tabular}}
\label{tab_5_gen}
\end{table}

\begin{sidewaysfigure}[]
\centering   
\includegraphics[width=19cm, scale=1]{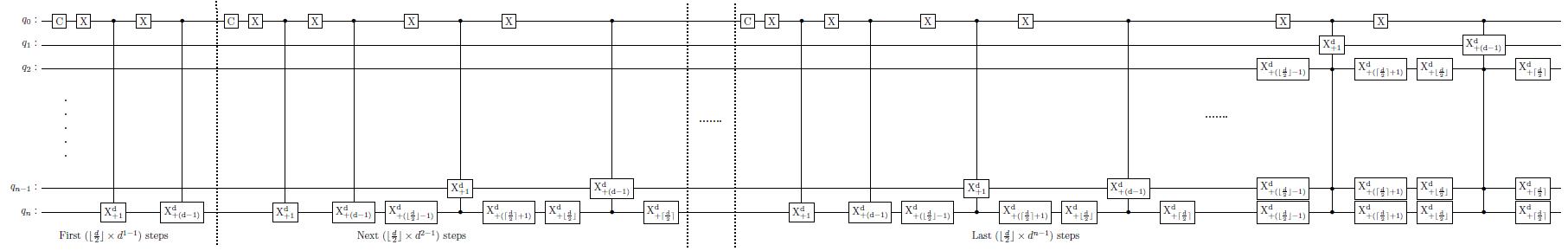}
    \caption{Generalized quantum circuit for the $\lfloor\frac{d^n}{2}\rfloor$ steps of DTQW on $n$ qudits in $d$-ary quantum systems when $d$ is odd.}
    \label{fig_gen_odd}
\end{sidewaysfigure}

As per the generalization of the proposed circuit, for a $n+1$-qubit-qudit $d$-ary system when $d$ is odd, after every $(\lfloor\frac{d}{2}\rfloor \times d^{q-1})$ (where $q$ is the number of qudits and $q$ ranges from 1 to $n$) steps, two new gates generalized multi-controlled Toffoli are added to realize the new position states, which is portrayed in Figure \ref{fig_gen_odd}. As an example, for a $n+1$-qubit-qudit 5-ary system, after every $(2 \times 5^{q-1})$  steps, two new generalized multi-controlled Toffoli gates are added to realize the new position states along with the previous set of gates (when $q>1$). In other words, if we follow Figure \ref{fig_gen_odd}, the first $(2 \times 5^{1-1})=2$ steps of DTQW in 5-ary systems have been implemented as per Figure \ref{fig_5_dtqw}. Next $(2 \times 5^{2-1})=10$ steps of DTQW in 5-ary systems have been implemented as per Figure \ref{fig_5_dtqw1}. Similarly, next $(2 \times 5^{3-1})=50$ steps of DTQW in 5-ary systems have been implemented as per Figure \ref{fig_5_dtqw2} and so on.

\subsection{ Quantum circuit for implementing the one-dimensional quantum walks in $d$-ary systems when $d$ is even}

{ We have already presented the logical realization of quantum circuits for implementing one-dimensional quantum walks using two-state coins in $d$-ary quantum systems when $d$ is odd. Now, we extend this approach to even values of $d$. Initially, we attempt to implement a one-dimensional (DTQW) using a nearest-neighbor approach similar to that used for odd dimensions. However, after a few steps, the walker reaches the same states in both directions, as shown in Table \ref{table_4_dtqw}, due to the even dimensionality. This symmetry prevents the walker from accessing the remaining state spaces. Thus, we observe that the direct nearest-neighbor mapping technique in even dimensions cannot be generalized for all $n$-qudit systems, as it fails to utilize the full state space. Consequently, we must rely on the naive increment-decrement approach \cite{ambainis2001one} to implement DTQW in even dimensions, ensuring that all state spaces in any finite-dimensional quantum system are utilized effectively.}

 \begin{table}[ht!]
\centering
\caption{{An example of position state mapping onto the multi-qudits states in 4-ary quantum systems using nearest-neighbor approach.}}
\begin{tabular}{|m{11em} | m{0.005em}| m{11em}  | }
\hline
~~~$\ket{x = 0} \equiv \ket{000}$  &&   \\
\hline
~~~$\ket{x = 1} \equiv \ket{001}$ && ~~~$\ket{x = -1} \equiv \ket{003}$\\
\hline 
~~~$\ket{x = 2} \equiv \ket{002}$  &&  ~~~$\ket{x = -2} \equiv \ket{032}$ \\
\hline
~~~$\ket{x = 3} \equiv \ket{013}$ && ~~~$\ket{x = -3} \equiv \ket{031}$\\
\hline
~~~$\ket{x = 4} \equiv \ket{010}$  && ~~~$\ket{x = -4} \equiv \ket{030}$\\
\hline
~~~$\ket{x = 5} \equiv \ket{011}$ &&  ~~~$\ket{x = -5} \equiv \ket{033}$\\
\hline 
~~~$\ket{x = 6} \equiv \ket{023}$ && ~~~$\ket{x = -6} \equiv \ket{021}$ \\
\hline
~~~$\ket{x = 7} \equiv \ket{020}$ && ~~~$\ket{x = -7} \equiv \ket{020}$ \\
\hline
\end{tabular}
\label{table_4_dtqw}
\end{table}

\begin{figure}[ht!]
  \centering 
\includegraphics[scale=0.6]{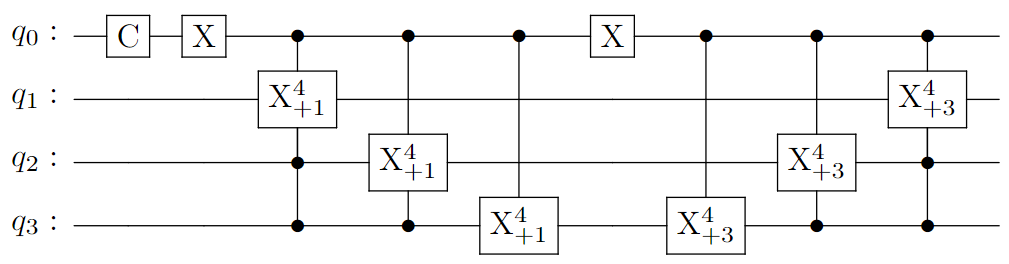}
    \caption{{Quantum circuit for the thirty-one steps of DTQW on three qudits in 4-ary quantum systems as shown in Table \ref{tab_4_dtqw}.}}
    \label{fig:41}
\end{figure}

 As an example, we have considered 4-ary and 6-ary quantum systems. For 4-ary quantum systems, to implement a DTQW in a one-dimensional position Hilbert space of size $4^q$, $(q)$ qudits of 4-ary quantum systems and one qubit are required, one qubit to represent the particle's internal state (coin qubit) and $q$- qudits to represent its position. For $q=3$, the number of steps of DTQW that can be implemented is $\lfloor(4^{q}-1)/2\rfloor = 31$. We choose the position state mapping given in Table \ref{tab_4_dtqw} with a fixed initial position state $\ket{000}$. A total of thirty-one DTQW steps have been presented in Table \ref{tab_4_dtqw}. Here also after each step of the DTQW, two new position states have to be considered. The corresponding quantum circuits for Table \ref{tab_4_dtqw} are portrayed in Figure \ref{fig:41}. It is to be noted that in the first step of DTQW, all three qudits need to be changed to describe the position states as shown in Table \ref{tab_4_dtqw}, unlike odd dimensions, for which only a change in one qudit was enough. For this reason, the implementation of DTQW in odd-dimensional qudit systems is more efficient as compared to the even-dimension.

\begin{table}[ht!]
\centering
\caption{An example of position state mapping onto the multi-qudits states in 4-ary quantum systems for quantum circuits presented in Figure \ref{fig:41}.}
\begin{tabular}{|m{11em} | m{0.005em}| m{11em}  | }
\hline
~~~$\ket{x = 0} \equiv \ket{000}$  &&   \\
\hline
~~~$\ket{x = 1} \equiv \ket{001}$ && ~~~$\ket{x = -1} \equiv \ket{333}$\\
\hline 
~~~$\ket{x = 2} \equiv \ket{002}$  &&  ~~~$\ket{x = -2} \equiv \ket{332}$ \\
\hline
~~~$\ket{x = 3} \equiv \ket{003}$ && ~~~$\ket{x = -3} \equiv \ket{331}$\\
\hline
~~~$\ket{x = 4} \equiv \ket{010}$  && ~~~$\ket{x = -4} \equiv \ket{330}$\\
\hline
~~~$\ket{x = 5} \equiv \ket{011}$ &&  ~~~$\ket{x = -5} \equiv \ket{323}$\\
\hline 
~~~$\ket{x = 6} \equiv \ket{012}$ && ~~~$\ket{x = -6} \equiv \ket{322}$ \\
\hline
~~~$\ket{x = 7} \equiv \ket{013}$ && ~~~$\ket{x = -7} \equiv \ket{321}$ \\
\hline
~~~$\ket{x = 8} \equiv \ket{020}$ && ~~~$\ket{x = -8} \equiv \ket{320}$ \\
\hline
~~~$\ket{x = 9} \equiv \ket{021}$ && ~~~$\ket{x = -9} \equiv \ket{313}$ \\
\hline
~~~$\ket{x = 10} \equiv \ket{022}$ && ~~~$\ket{x = -10} \equiv \ket{312}$ \\
\hline
~~~$\ket{x = 11} \equiv \ket{023}$ && ~~~$\ket{x = -11} \equiv \ket{311}$ \\
\hline
~~~$\ket{x = 12} \equiv \ket{030}$ && ~~~$\ket{x = -12} \equiv \ket{310}$ \\
\hline
~~~$\ket{x = 13} \equiv \ket{031}$ && ~~~$\ket{x = -13} \equiv \ket{303}$ \\
\hline
~~~$\ket{x = 14} \equiv \ket{032}$ && ~~~$\ket{x = -14} \equiv \ket{302}$\\
\hline 
~~~$\ket{x = 15} \equiv \ket{033}$  &&  ~~~$\ket{x = -15} \equiv \ket{301}$ \\
\hline
~~~$\ket{x = 16} \equiv \ket{100}$ && ~~~$\ket{x = -16} \equiv \ket{300}$\\
\hline
~~~$\ket{x = 17} \equiv \ket{101}$  && ~~~$\ket{x = -17} \equiv \ket{233}$\\
\hline
~~~$\ket{x = 18} \equiv \ket{102}$ && ~~~$\ket{x = -18} \equiv \ket{232}$\\
\hline
~~~$\ket{x = 19} \equiv \ket{103}$  && ~~~$\ket{x = -19} \equiv \ket{231}$\\
\hline
~~~$\ket{x = 20} \equiv \ket{110}$ &&  ~~~$\ket{x = -20} \equiv \ket{230}$\\
\hline 
~~~$\ket{x = 21} \equiv \ket{111}$ && ~~~$\ket{x = -21} \equiv \ket{223}$ \\
\hline
~~~$\ket{x = 22} \equiv \ket{112}$ && ~~~$\ket{x = -22} \equiv \ket{222}$ \\
\hline
~~~$\ket{x = 23} \equiv \ket{113}$ && ~~~$\ket{x = -23} \equiv \ket{221}$ \\
\hline
~~~$\ket{x = 24} \equiv \ket{120}$ && ~~~$\ket{x = -24} \equiv \ket{220}$ \\
\hline
~~~$\ket{x = 25} \equiv \ket{121}$ && ~~~$\ket{x = -25} \equiv \ket{213}$ \\
\hline
~~~$\ket{x = 26} \equiv \ket{122}$ && ~~~$\ket{x = -26} \equiv \ket{212}$ \\
\hline
~~~$\ket{x = 27} \equiv \ket{123}$ && ~~~$\ket{x = -27} \equiv \ket{211}$ \\
\hline
~~~$\ket{x = 28} \equiv \ket{130}$ && ~~~$\ket{x = -28} \equiv \ket{210}$ \\
\hline
~~~$\ket{x = 29} \equiv \ket{131}$ && ~~~$\ket{x = -29} \equiv \ket{203}$\\
\hline 
~~~$\ket{x = 30} \equiv \ket{132}$  &&  ~~~$\ket{x = -30} \equiv \ket{202}$ \\
\hline
~~~$\ket{x = 31} \equiv \ket{133}$ && ~~~$\ket{x = -31} \equiv \ket{201}$\\
\hline
\end{tabular}
\label{tab_4_dtqw}
\end{table}

{Again for a better understanding of position state mapping onto the multi-qudits states using an increment-decrement approach, we have presented another quantum system $i.e.,$ 6-ary quantum systems, which is thoroughly shown in Appendix \ref{6-ary}.}

\paragraph{Generalization of the quantum circuit for implementing one-dimensional discrete-time quantum walks in $d$-ary quantum systems when $d$ is even}

{The proposed circuits can also be scaled to implement more steps of one-dimensional DTQW on larger $d$-ary systems when $d$ is even with the help of higher controlled generalized MCT gates. As shown in Table \ref{tab_4_gen}, using $n+1$-qubit-qudit systems, implementation of $\lfloor (d^{n}-1)/2\rfloor$-steps of a DTQW can be performed. In other words, to implement the $n$-step of DTQW,  $\lceil\log_{d}(2n+1)\rceil$ qudits $+ 1$ qubit are required. Through mathematical induction, it can easily be proved that for $n$-step of DTQW,  $(\lceil\log_d{(2n+1)}\rceil)+1$ qubit-qudits are required, where $\lceil\log_d{(2n+1)}\rceil$ qudits for position states and $1$ qubit for the coin.}

\begin{table} [ht!]
\centering
\caption{Number of steps and a maximum number of control qudits required to control a target qubit in the DTQW for a 4-ary system of up to $n$ qudits.}
\scalebox{.8}{
\begin{tabular}{|m{6em}| m{6em} | m{6em} | m{7.5em} |}
\hline
No. of qudits & No. of qubit & No. of steps & Max. No. of controls in Generalized MCT gates\\
\hline
 1 & 1 & 1 & 1 \\
\hline
 2 & 1 & 7 & 2 \\
\hline 
 3 & 1 & 31 & 3\\
\hline
 4 & 1 & 127 & 4 \\
\hline
$n$ & 1 & $\lfloor (4^n-1)/2 \rfloor$ & n\\
\hline
\end{tabular}}
\label{tab_4_gen}
\end{table}

\begin{figure}[ht!]
    \centering
    \includegraphics[width=11cm, scale=1]{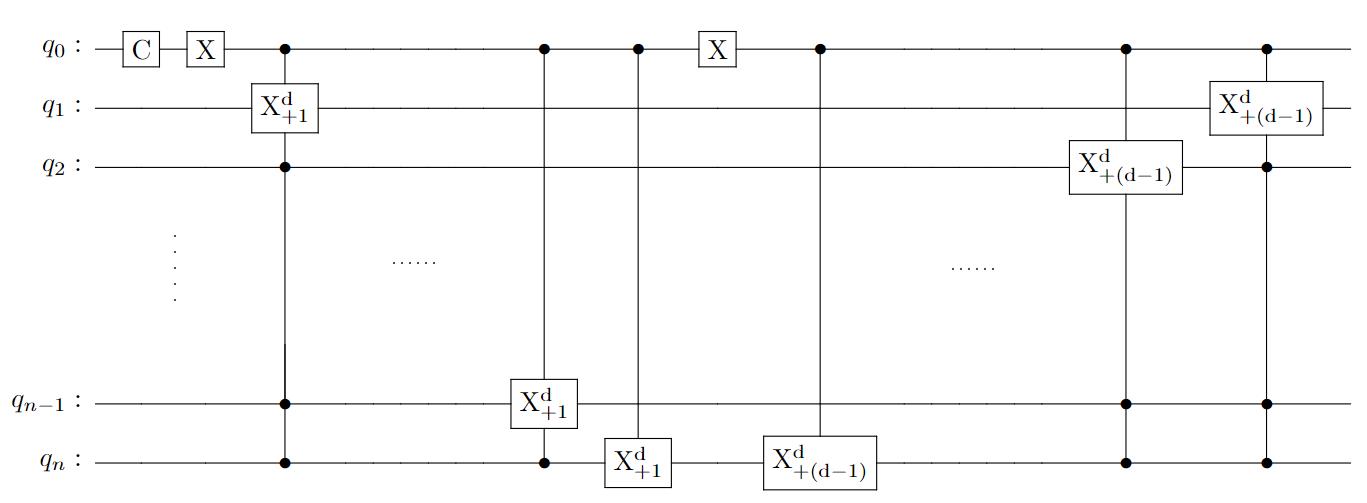}
    \caption{Quantum circuit for all the steps  of DTQW on $n$ qudits in $d$-ary quantum systems when $d$ is even.}
    \label{fig:deven1}
\end{figure}

As per the generalization of the proposed circuit for a $n+1$-qubit-qudit $d$-ary system, when $d$ is even, a quantum circuit is needed to be designed, which is shown in Figure \ref{fig:deven1}. From Figure \ref{fig_gen_odd} and \ref{fig:deven1}, it can be shown that from the first step of DTQW, all the values of qudits corresponding to the position states have been changed when $d$ is even, whether it takes a few DTQW steps to make a change on most significant qudit (MSQ) when $d$ is odd due to its nearest-neighbor approach. Therefore, we need $n$-controlled generalized MCT gates straightway from the first step of DTQW when $d$ is even, which makes the circuit costlier as compared to the design of the circuit when $d$ is odd.

In this way, we can also implement DTQW in two-dimensional position space, for this the same circuit can be scaled with an appropriate mapping of qudit states with the nearest-neighbor position space in both dimensions by introducing appropriate coin qubit/qudit into the $d$-ary systems. This approach can also be scaled up to $d$-dimensional position space. In such cases, the control over the target or position qudit increases with the number of coin qubits/qudits, which is thoroughly discussed in the next section.

\subsection{Quantum circuit for implementing quantum walks on higher dimensional lattices in an arbitrary finite-dimensional system}

In this section, the proposed quantum circuit has been designed in such a way, so that we can implement quantum walks on two or more dimensional lattices in an arbitrary finite-dimensional quantum system efficiently. 

\subsubsection{Quantum circuit for implementing two-dimensional DTQW on two-dimensional Lattices in an arbitrary finite-dimensional quantum System}
This section describes the proposed quantum circuit implementation of DTQW on a two-dimensional grid. Here, first, we present the logical realization of quantum circuits to implement two-dimensional quantum walks using four-dimensional coins in ternary quantum systems. To implement a DTQW in a two-dimensional position Hilbert space of size $3^q$, $\lfloor q/2\rfloor$ qutrits and two qubits are required, two qubits to represent the particle's internal state (coin qubit) and $\lfloor q/2\rfloor$-qutrits to represent its position. Similarly, the coin operation can be implemented by applying a single qubit rotation gate on the coin qubits. The position shift operation is subsequently implemented with the help of multi-qutrit gates where the coin qubits act as the control. Quantum circuits for implementing DTQW depend on how the position space is represented. Example circuits for a six-qubit-qutrit system are given in this section.

\begin{figure}[ht!]
\centering
\includegraphics[width=5cm, scale=1]{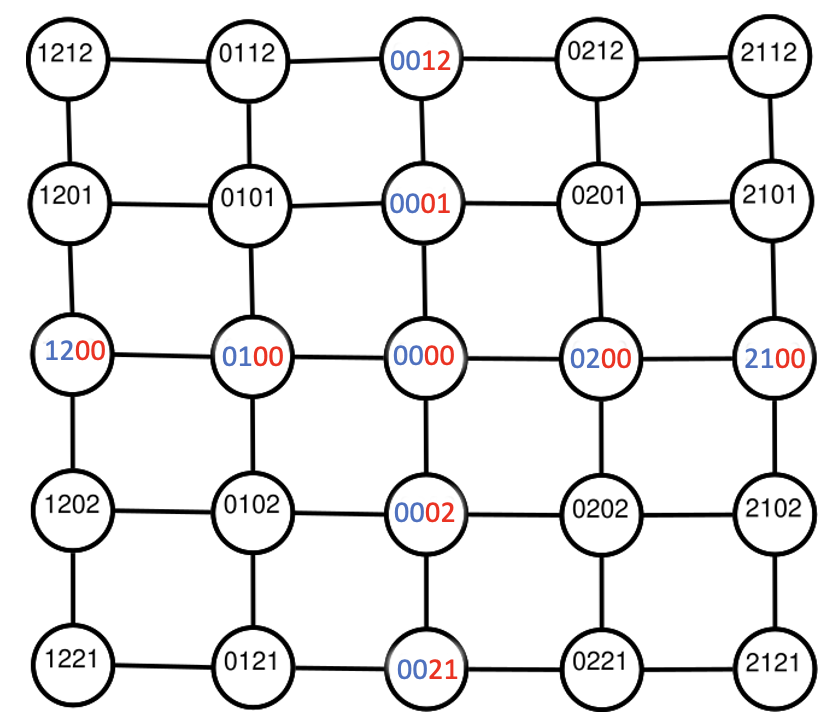}
\captionof{figure}{{An example of position state mapping onto the multi-qudits states in 3-ary quantum systems for quantum circuits presented in Figure \ref{fig:2d_31} and \ref{fig:2d_32}. The first two qutrits (color in blue) are considered for horizontal one-dimensional quantum walks, and the last two qutrits (color in red) are considered for vertical one-dimensional quantum walks.}}
\label{2d_3}
\end{figure}

For $q=4$, the number of steps of DTQW that can be implemented is $\lfloor\lfloor\sqrt{3^{q}}\rfloor/2\rfloor = 4$. We choose the position state mapping given in Figure \ref{2d_3} with a fixed initial position state $\ket{0000}$.  After each step of the DTQW, four new position states have to be considered. In Figure \ref{2d_3}, we show that the mapping of these new position states onto the multi-qutrit states is in such a way that the efficient number of gates are used to implement the shift operation, we consider the nearest-neighbor position space so as to make the circuit efficient. Each step of two-dimensional quantum walks can be considered as a combination of one-dimensional walks horizontally and one-dimensional walks vertically. Here, for each position state, the first two qutrits are considered for horizontal one-dimensional quantum walks, and the last two qutrits are considered for vertical one-dimensional quantum walks. As discussed, we have four coin states {00, 01, 10, 11} in which {00, 01} are used for {right shift, left shift} and {10, 11} are used for {down shift, up shift}. 

\begin{figure}[ht!]
\centering
\includegraphics[scale=0.5]{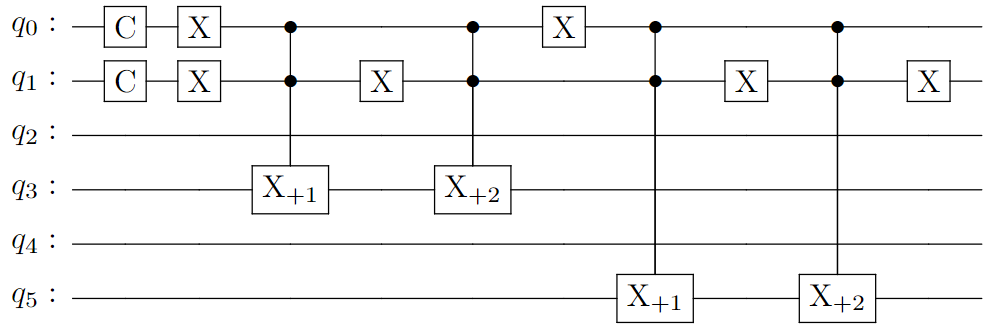}
    \caption{Quantum circuit for the first step of two-dimensional DTQW on four qutrits as shown in Figure \ref{2d_3}.}
    \label{fig:2d_31}
\end{figure}

\begin{figure*}[ht!]
\centering
\includegraphics[scale=0.4]{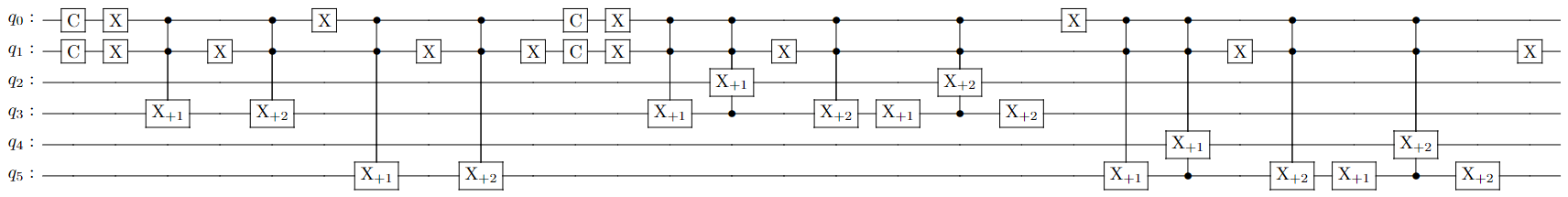}
    \caption{Quantum circuit for the second step of two-dimensional DTQW on four qutrits as shown in Figure \ref{2d_3}.}
    \label{fig:2d_32}
\end{figure*}

For the first step of two-dimensional DTQW as described in Figure \ref{2d_3}, the quantum circuit realization is shown in Figure \ref{fig:2d_31}. Figure \ref{fig:2d_32} portrays the quantum circuit realization for the second step of two-dimensional DTQW on four qutrits as shown in Figure \ref{2d_3}. {We have also considered another quantum system $i.e.,$ 5-ary quantum systems to illustrate the position state mapping onto the multi-qudits states for better understanding, which is thoroughly discussed in Appendix \ref{2d_appendix}.} In a similar way, two-dimensional DTQW can be implemented in an arbitrary finite-dimensional quantum system.

\subsubsection{Generalization of quantum circuit for implementing $d$-dimensional discrete-time quantum walks on $d$-dimensional lattices in an arbitrary finite-dimensional quantum system}

The proposed circuits can also be scaled to implement more steps of two-dimensional DTQW on larger $d$-ary systems with the help of higher controlled generalized MCT gates. Using $q+2$-qubit-qudit systems, implementation of $\lfloor\lfloor\sqrt{d^{q}}\rfloor/2\rfloor$-steps (where $q$ is always even since we have considered the same number of qudits for one-dimensional walks horizontally and one-dimensional walks vertically to perform the two-dimensional DTQW) of a two-dimensional DTQW can be performed. In other words, to implement $n$-step of two-dimensional DTQW,  $2 *\lceil\log_{d}n\rceil$ qudits $+ 2$ qubits are required.

\begin{figure}[ht!]
\centering
\includegraphics[width=10cm, scale=1]{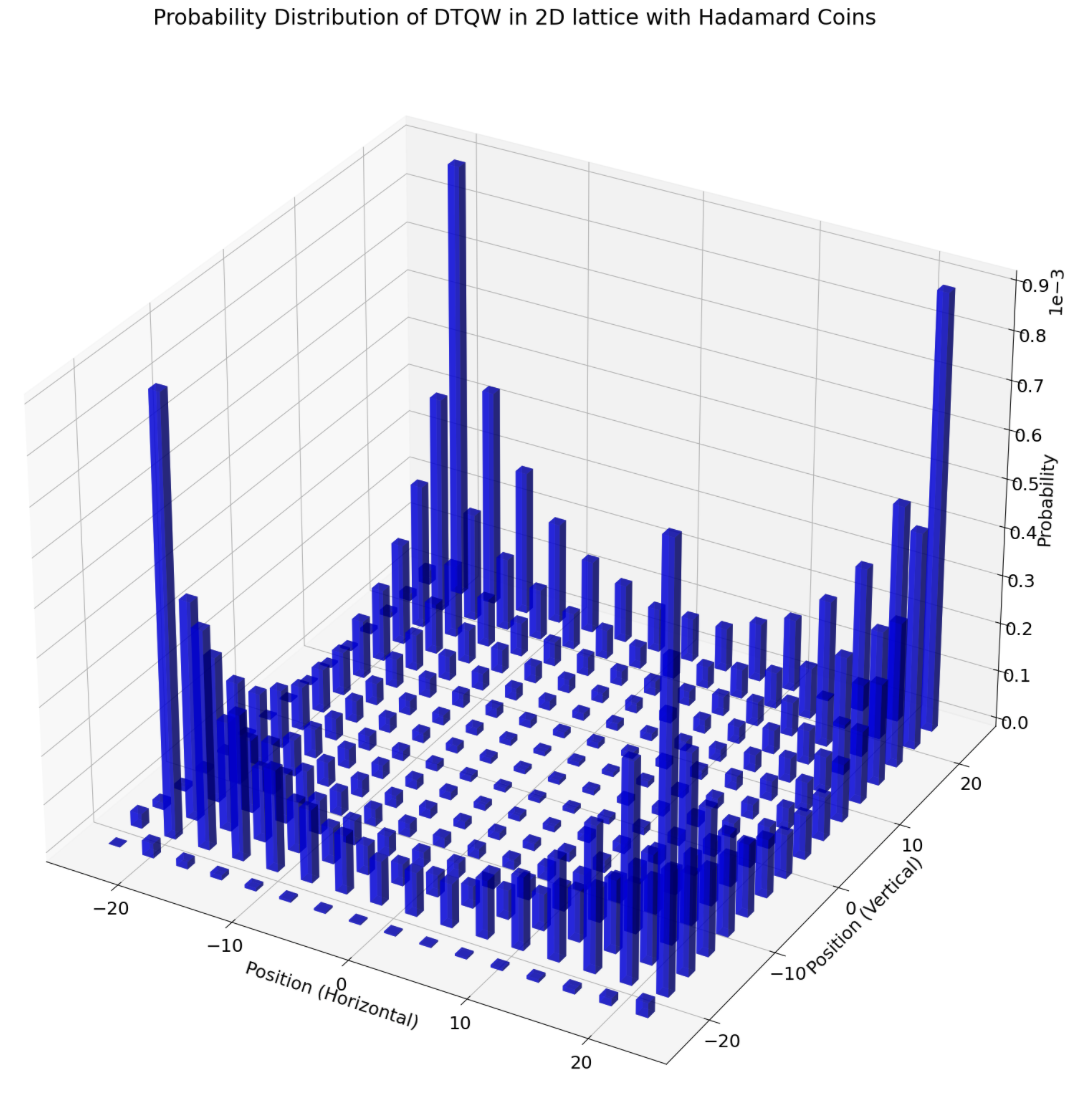}
\captionof{figure}{{Quantum walks on a two-dimensional lattice using a Hadamard coin, after 50 steps in 4-ary and 6-ary quantum systems. The X-axis and Y-axis denote position states and the Z-axis denotes the probability.}}
\label{H_dtqw_2d}
\end{figure}

Through numerical analysis, we have shown the probability distribution after 50 steps of a two-dimensional DTQW using a Hadamard coin for both the 4-ary and 6-ary quantum systems in Figure \ref{H_dtqw_2d}.

In this subsection, the concentration is on higher dimensional lattices. In this way, $d$ dimensional lattices can be realized. Our proposed circuits can also be scaled to implement $n$ steps of $d$-dimensional DTQW on larger $d$-ary systems with the help of higher controlled generalized MCT gates. As per the discussion in the previous subsection, using $q+d$-qubit-qudit systems, implementation of $\lfloor\lfloor\sqrt[d]{d^{q}}\rfloor/2\rfloor$-steps of a two-dimensional DTQW can be performed. In other words, to implement $n$-step of two-dimensional DTQW,  $d *\lceil\log_{d}n\rceil$ qudits $+ d$ qubits are required.


\section{Discussion}\label{4}
\paragraph{Alternative position state mapping onto the multi-qudits states}

\begin{figure}[ht!]
\centering   
\includegraphics[scale=0.5]{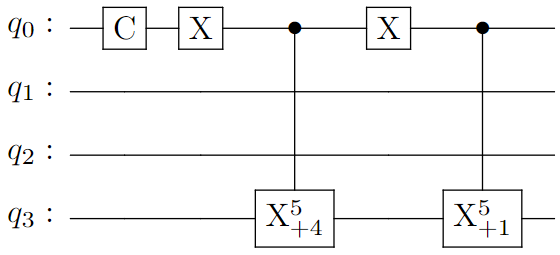}
    \caption{{Alternative} quantum circuit for the first two steps of DTQW on three qudits in 5-ary quantum systems as given in Table \ref{tab:alt5}.}
    \label{fig:alt51}
\end{figure}

\begin{figure}[ht!]
   \centering
\includegraphics[scale=.5]{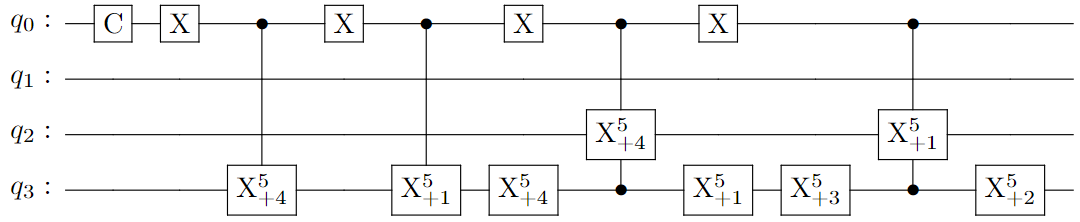}
    \caption{{Alternative} quantum circuit for the third-twelfth steps of DTQW on three qudits in 5-ary quantum systems as shown in Table \ref{tab:alt5}.}
    \label{fig:alt52}
\end{figure}

\begin{figure*}[ht!]
\centering   
\includegraphics[scale=0.42]{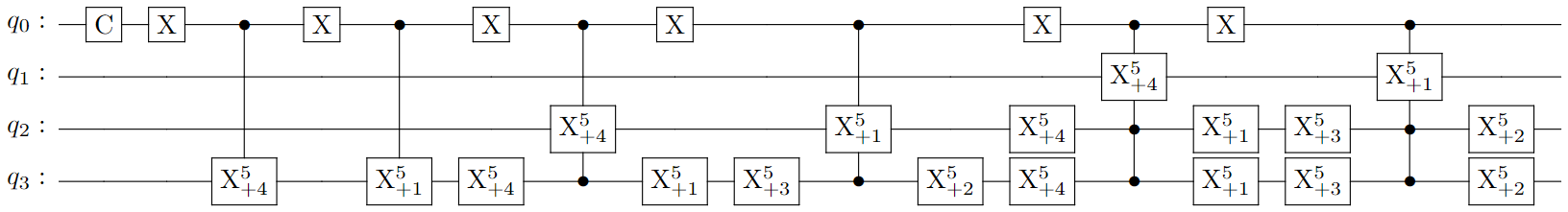}
    \caption{Alternative quantum circuit for the thirteenth-twenty fourth steps of DTQW on three qudits in 5-ary quantum systems as shown in Table \ref{tab:alt5}.}
    \label{fig:alt53}
\end{figure*}

The alternative quantum circuits for one-dimensional DTQW in 5-ary quantum systems are shown in Figure \ref{fig:alt51}, \ref{fig:alt52}, \ref{fig:alt53} for different mapping choice of position states (As we will always have two alternatives nearest-neighbor position spaces due to the different orthonormal basis states in qudit systems) onto multi-qudits states are shown in the Table \ref{tab:alt5}. These mapping choices are the only appropriate mapping of qudit states with the nearest-neighbor position space, making the circuit efficient and generalized for $n$ qudit systems. 


\begin{table}[ht!]
\centering
\caption{An alternative position state mapping onto the multi-qudits states in 5-ary quantum systems for quantum circuits presented in Figure \ref{fig:alt51}, Figure \ref{fig:alt52} and Figure \ref{fig:alt53}.}
\begin{tabular}{|m{11em} | m{0.005em}| m{11em}  | }
\hline
~~~ ~~~$\ket{x = 0} \equiv \ket{000}$  &&   \\
\hline
~~~  ~~~$\ket{x = 1} \equiv \ket{004}$ && $\ket{x = -1} \equiv \ket{001}$\\
\hline 
~~~  ~~~$\ket{x = 2} \equiv \ket{003}$ && $\ket{x = -2} \equiv \ket{002}$  \\
\hline
~~~  ~~~$\ket{x = 3} \equiv \ket{042}$ && $\ket{x = -3} \equiv \ket{013}$\\
\hline
~~~   ~~~$\ket{x = 4} \equiv \ket{041}$ && $\ket{x = -4} \equiv \ket{014}$\\
\hline
~~~ ~~~$\ket{x = 5} \equiv \ket{040}$ && $\ket{x = -5} \equiv \ket{010}$  \\
\hline 
~~~ ~~~$\ket{x = 6} \equiv \ket{044}$ &&  $\ket{x = -6} \equiv \ket{011}$ \\
\hline
~~~  ~~~$\ket{x = 7} \equiv \ket{043}$ &&  $\ket{x = -7} \equiv \ket{012}$\\
\hline
~~~  ~~~$\ket{x = 8} \equiv \ket{032}$ &&  $\ket{x = -8} \equiv \ket{023}$\\
\hline
~~~ ~~~$\ket{x = 9} \equiv \ket{031}$ &&  $\ket{x = -9} \equiv \ket{024}$ \\
\hline
~~~  ~~~$\ket{x = 10} \equiv \ket{030}$ && $\ket{x = -10} \equiv \ket{020}$\\
\hline
~~~ ~~~$\ket{x = 11} \equiv \ket{034}$ &&  $\ket{x = -11} \equiv \ket{021}$\\
\hline
~~~  ~~~$\ket{x = 12} \equiv \ket{033}$ && $\ket{x = -12} \equiv \ket{022}$\\
\hline
~~~  ~~~$\ket{x = 13} \equiv \ket{422}$ &&  $\ket{x = -13} \equiv \ket{133}$\\

\hline
~~~ ~~~$\ket{x = 14} \equiv \ket{421}$  &&  $\ket{x = -14} \equiv \ket{134}$ \\
\hline 
~~~   ~~~$\ket{x = 15} \equiv \ket{420}$ && $\ket{x = -15} \equiv \ket{130}$ \\
\hline
~~~  ~~~$\ket{x = 16} \equiv \ket{424}$ &&  $\ket{x = -16} \equiv \ket{131}$\\
\hline
~~~   ~~~$\ket{x = 17} \equiv \ket{423}$  &&  $\ket{x = -17} \equiv \ket{132}$\\
\hline
~~~ ~~~$\ket{x = 18} \equiv \ket{412}$ && $\ket{x = -18} \equiv \ket{143}$\\
\hline 
~~~ ~~~$\ket{x = 19} \equiv \ket{411}$ &&  $\ket{x = -19} \equiv \ket{144}$ \\
\hline
~~~  ~~~$\ket{x = 20} \equiv \ket{410}$ &&  $\ket{x = -20} \equiv \ket{140}$\\
\hline
~~~ ~~~$\ket{x = 21} \equiv \ket{414}$ &&  $\ket{x = -21} \equiv \ket{141}$ \\
\hline
~~~  ~~~$\ket{x = 22} \equiv \ket{413}$ &&  $\ket{x = -22} \equiv \ket{142}$\\
\hline
~~~  ~~~$\ket{x = 23} \equiv \ket{402}$ &&  $\ket{x = -23} \equiv \ket{103}$\\
\hline
~~~ ~~~$\ket{x = 24} \equiv \ket{401}$ && $\ket{x = -24} \equiv \ket{104}$ \\
\hline
\end{tabular}
\label{tab:alt5}
\end{table}

\begin{figure*}[ht!]
\centering
\includegraphics[width=11.8cm, scale=1]{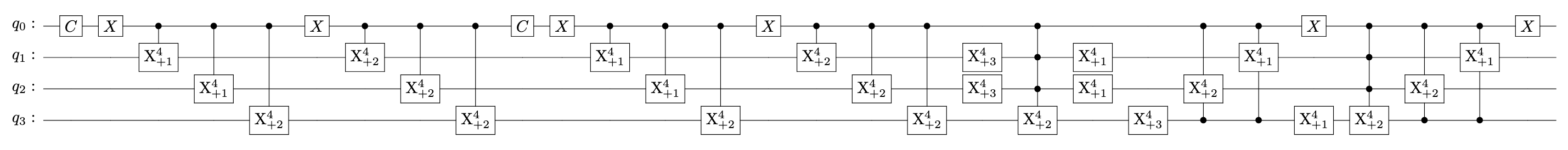}
\captionof{figure}{{Quantum circuit for first two steps of the DTQW on a four ququad system with a fixed initial state $\ket{\uparrow} \otimes \ket{x=0} \equiv \ket{\uparrow} \otimes \ket{000}$. The position state mapping is shown in Table\,\ref{example}.}}
\label{examplepic}
\end{figure*}

\paragraph{Efficiency analysis of proposed position state mapping onto the multi-qudit states} 
As discussed, these appropriate position state mapping using the nearest-neighbor approach make our implementation efficient in terms of gate cost and the number of qudits, when the dimension of qudit systems is odd. Albeit, from the first step of DTQW, all the values of qudits corresponding to the position states have been changed when $d$ is even since it does not follow the nearest-neighbor approach, rather follows increment-decrement approach, therefore, the circuit is more costlier as compared to the design of the circuit when $d$ is odd. As an example in Table \ref{tab_4_dtqw}, we illustrate that the LSQ in 5-ary quantum systems changes in every step of quantum walks by following the nearest-neighbor approach. But the next qudit towards the left of LSQ and the MSQ changes for the minimum possible time by following the nearest-neighbor approach. The gates in the quantum circuit are required for the change in qudit states. Hence, the gate count in each step of quantum walks can also be minimized. Apart from these mapping choices, any naive mapping choices including increment-decrement approach of the position state onto the qudit states will lead to an inefficient quantum circuit with a higher number of quantum gates as they don't follow nearest-neighbor logic. One such example is given in Table \ref{example} and Figure \ref{examplepic}. In this example, due to the configuration of the mapped position state, only two steps of DTQW can be performed, which is shown in Figure \ref{examplepic} whereas, using our position state mapping approach, $31$ steps of DTQW can be realized in the same system. Thus the naive mapping choices of the position states-based circuit consist of many additional qudit gates compared to the nearest-neighbor position space-based circuits as shown in Figure \ref{fig:41}. The next subsection will discuss the comparative analysis of different quantum systems.

\begin{table}[ht!]
\centering
\caption{An example of mapping of position state onto the multi-qudits states for quantum circuits presented in Figure \ref{examplepic}.}
\begin{tabular}{|m{11em} | m{0.005em}| m{11em}  | }
\hline
~~~$\ket{x = 0} \equiv \ket{000}$  &&   \\
\hline
~~~$\ket{x = 1} \equiv \ket{222}$ && ~~~$\ket{x = -1} \equiv \ket{112}$\\
\hline 
~~~$\ket{x = 2} \equiv \ket{021}$  &&  ~~~$\ket{x = -2} \equiv \ket{012}$ \\
\hline
~~~$\ket{x = 3} \equiv \ket{212}$ && ~~~$\ket{x = -3} \equiv \ket{211}$\\
\hline
\end{tabular}
\label{example}
\end{table}

\begin{figure}[ht!]
\centering
\includegraphics[width=6cm, scale=1]{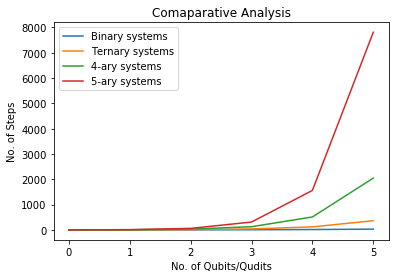}
\captionof{figure}{Comparative analysis of DTQW for different quantum systems.}
\label{qvs}
\end{figure}

\paragraph{Comparative analysis of DTQW for different quantum systems} 
Figure \ref{qvs} shows the variation in the number of steps for the binary, ternary, 4-ary, and 5-ary quantum systems and it clearly shows that the 5-ary quantum systems outperform the binary, ternary, and 4-ary quantum systems quite comprehensibly.

\begin{figure}[!h]
            \centering
            \includegraphics[scale=.2]{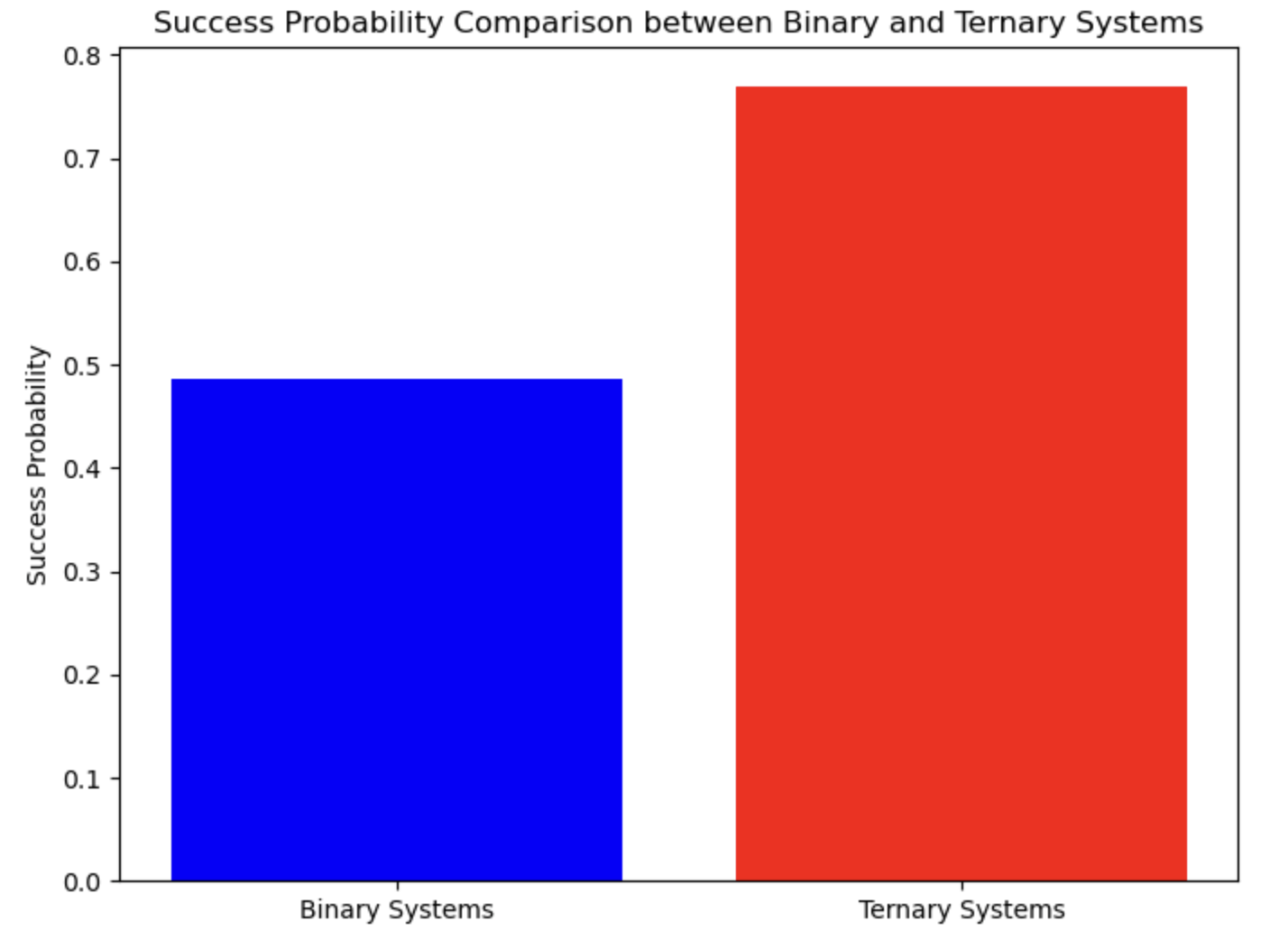}
                \caption{{Success probability analysis between 3-qubit binary systems and 3-qutrit ternary systems for 4-step DTQW.}}
                \label{fig:success_prob_analysis_vs_naive}
            \end{figure}

{To analyze the noise of the proposed circuit on a superconducting processor, first, we need to calculate the depth and the gate count of the circuit. The circuit depth of a quantum circuit refers to the number of layers of quantum gates that can be executed simultaneously to complete the entire computation. In the proposed circuit, we see a heavy use of multi-controlled Toffoli gates. As the number of qubits or qudits in the circuit increases, the number of these multi-controlled Toffoli gates grows significantly. However, these gates cannot be directly executed on existing quantum hardware and must first be decomposed into single- and two-qubit (or qudit) gates. The gate count refers to the total number of gates in the circuit after this decomposition process. Intermediate qudits can also be used to decompose Toffoli gates, as proposed by the authors in \cite{saha2020asymptotically, gokhale2019asymptotic} for ancilla-free decomposition. When multi-controlled Toffoli gates are decomposed, the circuit's depth and gate count will inevitably increase, as each gate is replaced by multiple others. Our goal is to maximize the probability of success for these circuits. Quantum gates are prone to small errors, which can be modeled as an ideal gate followed by an undesired Pauli operator. Rather than focusing on the likelihood of small errors in the circuit, we compare the probability that the circuit remains error-free (the success probability), while maintaining the same level of generality as in \cite{saha2020asymptotically}. Assuming each component fails independently, the generalized formula for the success probability ($P_{success}$) for any decomposition is given by the product of individual success probabilities. The general equation for calculating the success probability is,
            \begin{equation}\label{prob_success}
                P_{success}=\prod_{gates}{((P_{\text{success of gate}})^{\#gates}\times e^{-depth/T_1})}
            \end{equation}
           where $P_{success}$ represents the success probability of the decomposed circuit, $P_{\text{success of gate}}$ is the success probability of each gate (whether single-qubit/qudit or multi-qubit/qudit) based on the hardware specifications, $depth$ refers to the circuit depth after decomposition and $T_1$ is the relaxation time of the quantum hardware. For further experiments, we take the value of $T_1$ as per \cite{IBMnoise} for different available quantum systems up to 4-ary quantum systems. Hence, we need to restrict to smaller dimensional quantum systems for further comparative analysis. We take 3-qubit binary systems and 3-qutrit ternary systems for 4-step DTQW. The probability of success for both the systems is as Figure \ref{fig:success_prob_analysis_vs_naive}. It clearly exhibits that instead of a higher error rate in higher dimensional systems, the nearest-neighbor approach on ternary systems outperforms the increment-decrement approach on binary systems with respect to success probability for the same number of DTQW steps due to robust circuit concerning circuit depth and gate count.}

\paragraph{Implementation of DTQW using different coins}

\begin{figure}[ht!]
\centering
\includegraphics[width=6cm, scale=1]{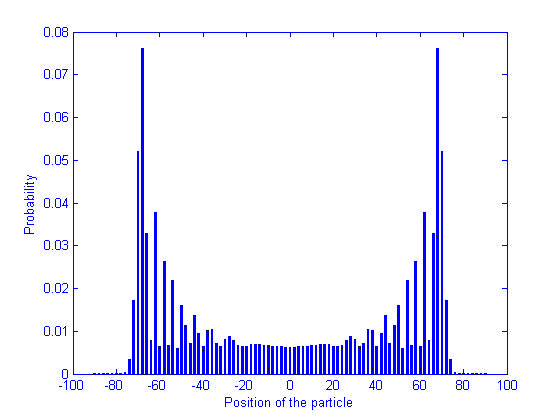}
\captionof{figure}{{Quantum walks on a one-dimensional line using Hadamard coin, after 100 steps \cite{9410395}.}}
\label{H_dtqw}
\end{figure}

\begin{figure}[ht!]
\centering
\includegraphics[width=5cm, scale=1]{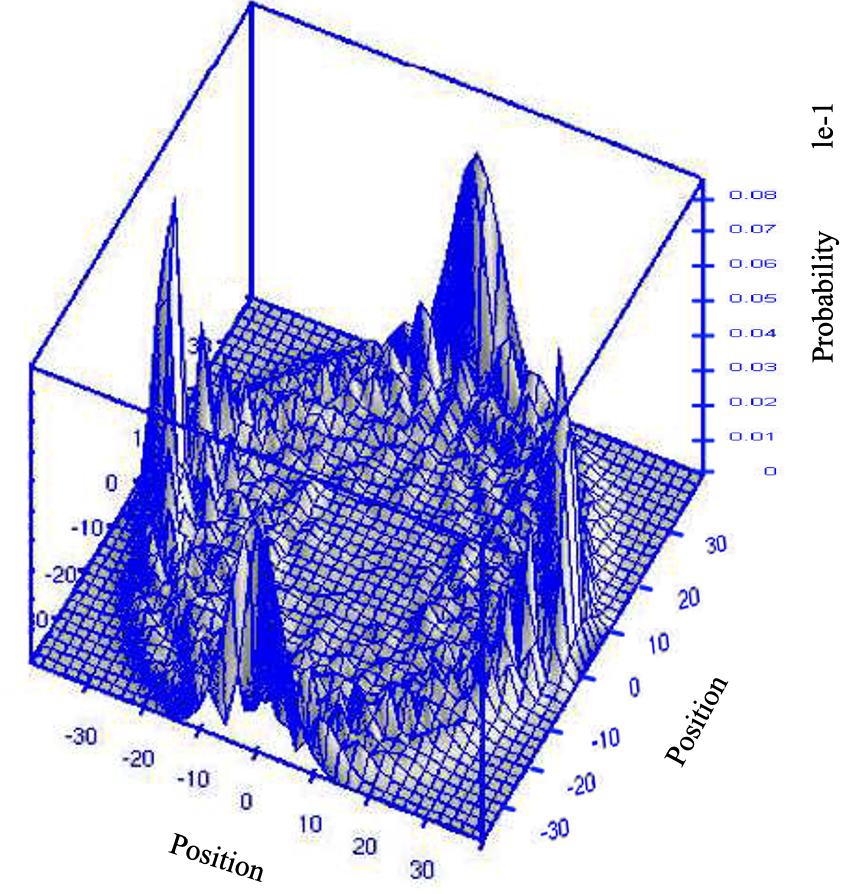}
\captionof{figure}{{Two-dimensional discrete-time quantum walks on a two-dimensional lattice using a DFT coin, after 50 steps. The X-axis and Y-axis denote position state and the Z-axis denotes the probability.}}
\label{2d_DFT}
\end{figure}

\begin{figure}[ht!]
\centering
\includegraphics[width=5cm, scale=1]{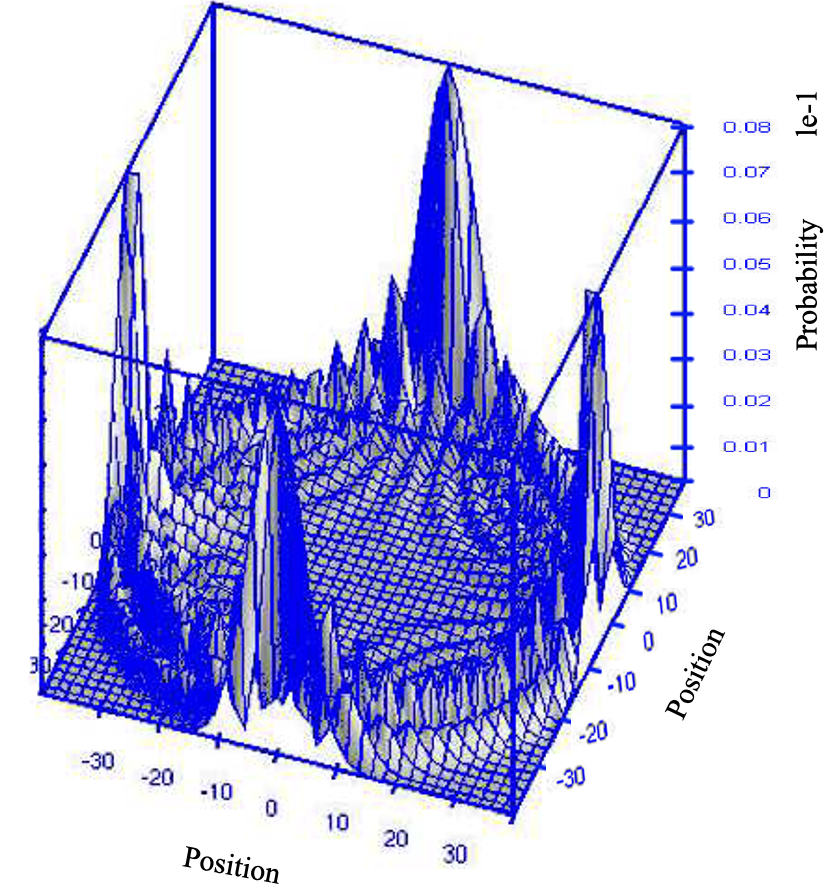}
\captionof{figure}{{Two-dimensional discrete-time quantum walks on a two-dimensional lattice using Grover's coin, after 50 steps. The X-axis and Y-axis denote position state and the Z-axis denotes the probability.}}
\label{2d_grover}
\end{figure}

Through numerical analysis, we have simulated and verified our proposed circuit designs of DTQW. Figure \ref{H_dtqw} shows the probability distribution after 100 steps of a DTQW using a Hadamard coin, where the initial position state is $\ket{0}$. Figure \ref{2d_DFT} and \ref{2d_grover} portray the probability distribution after 50 steps of a two-dimensional DTQW using a DFT coin and Grover's coin respectively, where the initial position state is as per \cite{Tregenna_2003}. The diffusiveness of the probability distribution of DTQW can be shown from the results, which differentiates the DTQW from classical random walks. 

\section{Conclusion}\label{5}

In this work, we have defined one-dimensional DTQW using two-dimensional coins in $d$-dimensional quantum systems, where $d>2$. Further, we have proposed quantum circuit realization to implement one-dimensional DTQW in qudit systems using an appropriate logical mapping of the position space on which a walker evolves onto the multi-qudit states. We also showed the implementation of one-dimensional DTQW in odd-dimensional qudit systems is much more efficient as compared to the even-dimension. Later, we also address the scalability of the proposed circuit to $n$-qudit systems. Further, we have exhibited an efficient quantum circuit realization to implement $d$-dimensional DTQW in $d$-dimensional quantum lattice with an example circuit for two-dimensional DTQW in a 2D grid for the first of its kind using an appropriate logical mapping of the position space on which a walker evolves onto the multi-qudit states. Finally, we efficiently implemented DTQW for different coin operators considering various position-space. We have also verified our proposed circuits through simulation. In this article, we achieved the implementation of the maximum possible number of steps of any finite-dimensional DTQW in any finite-dimensional qudit systems with the minimum possible number of qubits/qudits through our proposed approach. This efficient DTQW state preparation will provide direct efficacy to implement various applications such as communication protocols \cite{Srikara_2020} or searching problems \cite{Saha_2022}. In the future, we will try to reduce the number of gates to implement DTQW by adding additional ancilla qudits. This will help one to implement more steps of the DTQW keeping the circuit depth constant. We shall also try to find a generic solution for implementation of DTQW using nearest-neighbor approach in future. The results are very promising to pave the way for further research in qudit-assisted quantum computing. With the evolution of qudit-supported quantum hardware, we would like to validate our designs in the near future.
\backmatter

\section*{Declarations}

\begin{itemize}
\item Availability of data and materials: Not applicable.
\item Conflict of interest: There is no conflict of interest.
\end{itemize}

\bibliographystyle{sn-mathphys}
\bibliography{sn-bibliography}


\begin{thebibliography}{51}
\ifx \bisbn   \undefined \def \bisbn  #1{ISBN #1}\fi
\ifx \binits  \undefined \def \binits#1{#1}\fi
\ifx \bauthor  \undefined \def \bauthor#1{#1}\fi
\ifx \batitle  \undefined \def \batitle#1{#1}\fi
\ifx \bjtitle  \undefined \def \bjtitle#1{#1}\fi
\ifx \bvolume  \undefined \def \bvolume#1{\textbf{#1}}\fi
\ifx \byear  \undefined \def \byear#1{#1}\fi
\ifx \bissue  \undefined \def \bissue#1{#1}\fi
\ifx \bfpage  \undefined \def \bfpage#1{#1}\fi
\ifx \blpage  \undefined \def \blpage #1{#1}\fi
\ifx \burl  \undefined \def \burl#1{\textsf{#1}}\fi
\ifx \doiurl  \undefined \def \doiurl#1{\url{https://doi.org/#1}}\fi
\ifx \betal  \undefined \def \betal{\textit{et al.}}\fi
\ifx \binstitute  \undefined \def \binstitute#1{#1}\fi
\ifx \binstitutionaled  \undefined \def \binstitutionaled#1{#1}\fi
\ifx \bctitle  \undefined \def \bctitle#1{#1}\fi
\ifx \beditor  \undefined \def \beditor#1{#1}\fi
\ifx \bpublisher  \undefined \def \bpublisher#1{#1}\fi
\ifx \bbtitle  \undefined \def \bbtitle#1{#1}\fi
\ifx \bedition  \undefined \def \bedition#1{#1}\fi
\ifx \bseriesno  \undefined \def \bseriesno#1{#1}\fi
\ifx \blocation  \undefined \def \blocation#1{#1}\fi
\ifx \bsertitle  \undefined \def \bsertitle#1{#1}\fi
\ifx \bsnm \undefined \def \bsnm#1{#1}\fi
\ifx \bsuffix \undefined \def \bsuffix#1{#1}\fi
\ifx \bparticle \undefined \def \bparticle#1{#1}\fi
\ifx \barticle \undefined \def \barticle#1{#1}\fi
\bibcommenthead
\ifx \bconfdate \undefined \def \bconfdate #1{#1}\fi
\ifx \botherref \undefined \def \botherref #1{#1}\fi
\ifx \url \undefined \def \url#1{\textsf{#1}}\fi
\ifx \bchapter \undefined \def \bchapter#1{#1}\fi
\ifx \bbook \undefined \def \bbook#1{#1}\fi
\ifx \bcomment \undefined \def \bcomment#1{#1}\fi
\ifx \oauthor \undefined \def \oauthor#1{#1}\fi
\ifx \citeauthoryear \undefined \def \citeauthoryear#1{#1}\fi
\ifx \endbibitem  \undefined \def \endbibitem {}\fi
\ifx \bconflocation  \undefined \def \bconflocation#1{#1}\fi
\ifx \arxivurl  \undefined \def \arxivurl#1{\textsf{#1}}\fi
\csname PreBibitemsHook\endcsname

\bibitem{nielsen_chuang_2010}
\begin{bbook}
\bauthor{\bsnm{Nielsen}, \binits{M.A.}},
\bauthor{\bsnm{Chuang}, \binits{I.L.}}:
\bbtitle{Quantum Computation and Quantum Information: 10th Anniversary Edition},
\bedition{10th} edn.
\bpublisher{Cambridge University Press},
\blocation{USA}
(\byear{2011})
\end{bbook}
\endbibitem

\bibitem{Farhi_1998}
\begin{barticle}
\bauthor{\bsnm{Farhi}, \binits{E.}},
\bauthor{\bsnm{Gutmann}, \binits{S.}}:
\batitle{Quantum computation and decision trees}.
\bjtitle{Physical Review A}
\bvolume{58}(\bissue{2}),
\bfpage{915}--\blpage{928}
(\byear{1998}).
\doiurl{10.1103/physreva.58.915}
\end{barticle}
\endbibitem

\bibitem{PhysRevA.48.1687}
\begin{barticle}
\bauthor{\bsnm{Aharonov}, \binits{Y.}},
\bauthor{\bsnm{Davidovich}, \binits{L.}},
\bauthor{\bsnm{Zagury}, \binits{N.}}:
\batitle{Quantum random walks}.
\bjtitle{Phys. Rev. A}
\bvolume{48},
\bfpage{1687}--\blpage{1690}
(\byear{1993}).
\doiurl{10.1103/PhysRevA.48.1687}
\end{barticle}
\endbibitem

\bibitem{Shenvi_2003}
\begin{botherref}
\oauthor{\bsnm{Shenvi}, \binits{N.}},
\oauthor{\bsnm{Kempe}, \binits{J.}},
\oauthor{\bsnm{Whaley}, \binits{K.B.}}:
Quantum random-walk search algorithm.
Physical Review A
\textbf{67}(5)
(2003).
\doiurl{10.1103/physreva.67.052307}
\end{botherref}
\endbibitem

\bibitem{ambainis2004quantum}
\begin{botherref}
\oauthor{\bsnm{Ambainis}, \binits{A.}}:
Quantum walks and their algorithmic applications
(2004)
\end{botherref}
\endbibitem

\bibitem{aharonov2000quantum}
\begin{botherref}
\oauthor{\bsnm{Aharonov}, \binits{D.}},
\oauthor{\bsnm{Ambainis}, \binits{A.}},
\oauthor{\bsnm{Kempe}, \binits{J.}},
\oauthor{\bsnm{Vazirani}, \binits{U.}}:
Quantum Walks On Graphs
(2000)
\end{botherref}
\endbibitem

\bibitem{magniez2003quantum}
\begin{botherref}
\oauthor{\bsnm{Magniez}, \binits{F.}},
\oauthor{\bsnm{Santha}, \binits{M.}},
\oauthor{\bsnm{Szegedy}, \binits{M.}}:
Quantum Algorithms for the Triangle Problem
(2003)
\end{botherref}
\endbibitem

\bibitem{ambainis2003quantum}
\begin{botherref}
\oauthor{\bsnm{Ambainis}, \binits{A.}}:
Quantum walk algorithm for element distinctness
(2003)
\end{botherref}
\endbibitem

\bibitem{Tregenna_2003}
\begin{barticle}
\bauthor{\bsnm{Tregenna}, \binits{B.}},
\bauthor{\bsnm{Flanagan}, \binits{W.}},
\bauthor{\bsnm{Maile}, \binits{R.}},
\bauthor{\bsnm{Kendon}, \binits{V.}}:
\batitle{Controlling discrete quantum walks: coins and initial states}.
\bjtitle{New Journal of Physics}
\bvolume{5}(\bissue{1}),
\bfpage{83}
(\byear{2003}).
\doiurl{10.1088/1367-2630/5/1/383}
\end{barticle}
\endbibitem

\bibitem{Childs_2003}
\begin{barticle}
\bauthor{\bsnm{Childs}, \binits{A.M.}},
\bauthor{\bsnm{Cleve}, \binits{R.}},
\bauthor{\bsnm{Deotto}, \binits{E.}},
\bauthor{\bsnm{Farhi}, \binits{E.}},
\bauthor{\bsnm{Gutmann}, \binits{S.}},
\bauthor{\bsnm{Spielman}, \binits{D.A.}}:
\batitle{Exponential algorithmic speedup by a quantum walk}.
\bjtitle{Proceedings of the thirty-fifth ACM symposium on Theory of computing - STOC ’03}
(\byear{2003}).
\doiurl{10.1145/780542.780552}
\end{barticle}
\endbibitem

\bibitem{Kempe03}
\begin{barticle}
\bauthor{\bsnm{Kempe}, \binits{J.}}:
\batitle{Quantum random walks: An introductory overview}.
\bjtitle{Contemporary Physics}
\bvolume{44}(\bissue{4}),
\bfpage{307}--\blpage{327}
(\byear{2003}).
\doiurl{10.1080/00107151031000110776}
\end{barticle}
\endbibitem

\bibitem{Venegas_Andraca_2012}
\begin{barticle}
\bauthor{\bsnm{Venegas-Andraca}, \binits{S.E.}}:
\batitle{Quantum walks: a comprehensive review}.
\bjtitle{Quantum Information Processing}
\bvolume{11}(\bissue{5}),
\bfpage{1015}--\blpage{1106}
(\byear{2012}).
\doiurl{10.1007/s11128-012-0432-5}
\end{barticle}
\endbibitem

\bibitem{Kendon_2006}
\begin{barticle}
\bauthor{\bsnm{Kendon}, \binits{V.M.}}:
\batitle{A random walk approach to quantum algorithms}.
\bjtitle{Philosophical Transactions of the Royal Society A: Mathematical, Physical and Engineering Sciences}
\bvolume{364}(\bissue{1849}),
\bfpage{3407}--\blpage{3422}
(\byear{2006}).
\doiurl{10.1098/rsta.2006.1901}
\end{barticle}
\endbibitem

\bibitem{32}
\begin{bchapter}
\bauthor{\bsnm{Grover}, \binits{L.K.}}:
\bctitle{A fast quantum mechanical algorithm for database search}.
In: \bbtitle{Proceedings of the Twenty-eighth Annual ACM Symposium on Theory of Computing}.
\bsertitle{STOC '96},
pp. \bfpage{212}--\blpage{219}.
\bpublisher{ACM},
\blocation{New York, NY, USA}
(\byear{1996}).
\doiurl{10.1145/237814.237866}
\end{bchapter}
\endbibitem

\bibitem{31}
\begin{barticle}
\bauthor{\bsnm{Shor}, \binits{P.W.}}:
\batitle{Polynomial-time algorithms for prime factorization and discrete logarithms on a quantum computer}.
\bjtitle{SIAM Journal on Computing}
\bvolume{26}(\bissue{5}),
\bfpage{1484}--\blpage{1509}
(\byear{1997}).
\doiurl{10.1137/s0097539795293172}
\end{barticle}
\endbibitem

\bibitem{Ringbauer_2022}
\begin{barticle}
\bauthor{\bsnm{Ringbauer}, \binits{M.}},
\bauthor{\bsnm{Meth}, \binits{M.}},
\bauthor{\bsnm{Postler}, \binits{L.}},
\bauthor{\bsnm{Stricker}, \binits{R.}},
\bauthor{\bsnm{Blatt}, \binits{R.}},
\bauthor{\bsnm{Schindler}, \binits{P.}},
\bauthor{\bsnm{Monz}, \binits{T.}}:
\batitle{A universal qudit quantum processor with trapped ions}.
\bjtitle{Nature Physics}
\bvolume{18}(\bissue{9}),
\bfpage{1053}--\blpage{1057}
(\byear{2022}).
\doiurl{10.1038/s41567-022-01658-0}
\end{barticle}
\endbibitem

\bibitem{1}
\begin{barticle}
\bauthor{\bsnm{Brouwer}, \binits{A.E.}},
\bauthor{\bsnm{Shearer}, \binits{J.B.}},
\bauthor{\bsnm{Sloane}, \binits{N.J.A.}},
\bauthor{\bsnm{Smith}, \binits{W.D.}}:
\batitle{A new table of constant weight codes}.
\bjtitle{IEEE Transactions on Information Theory}
\bvolume{36}(\bissue{6}),
\bfpage{1334}--\blpage{1380}
(\byear{1990}).
\doiurl{10.1109/18.59932}
\end{barticle}
\endbibitem

\bibitem{6}
\begin{botherref}
\oauthor{\bsnm{Matsunaga}, \binits{T.}},
\oauthor{\bsnm{Yonemori}, \binits{C.}},
\oauthor{\bsnm{Tomita}, \binits{E.}}:
Clique-based data mining for related genes in a biomedical database.
BMC Bioinformatics
\textbf{10}(205)
(2009).
\doiurl{10.1186/1471-2105-10-205}
\end{botherref}
\endbibitem

\bibitem{13}
\begin{bchapter}
\bauthor{\bsnm{Ambainis}, \binits{A.}}:
\bctitle{Quantum walk algorithm for element distinctness}.
In: \bbtitle{45th Annual IEEE Symposium on Foundations of Computer Science},
pp. \bfpage{22}--\blpage{31}
(\byear{2004}).
\doiurl{10.1109/FOCS.2004.54}
\end{bchapter}
\endbibitem

\bibitem{17}
\begin{barticle}
\bauthor{\bsnm{Aghaei}, \binits{M.}},
\bauthor{\bsnm{Zukarnain}, \binits{Z.}},
\bauthor{\bsnm{Mamat}, \binits{A.}},
\bauthor{\bsnm{Zainuddin}, \binits{H.}}:
\batitle{A hybrid architecture approach for quantum algorithms.}
\bjtitle{Journal of Computer Science}
\bvolume{5},
\bfpage{725}--\blpage{731}
(\byear{2009})
\end{barticle}
\endbibitem

\bibitem{19}
\begin{barticle}
\bauthor{\bsnm{Bron}, \binits{C.}},
\bauthor{\bsnm{Kerbosch}, \binits{J.}}:
\batitle{Algorithm 457: Finding all cliques of an undirected graph}.
\bjtitle{Commun. ACM}
\bvolume{16}(\bissue{9}),
\bfpage{575}--\blpage{577}
(\byear{1973}).
\doiurl{10.1145/362342.362367}
\end{barticle}
\endbibitem

\bibitem{34}
\begin{bchapter}
\bauthor{\bsnm{Buhrman}, \binits{H.}},
\bauthor{\bsnm{\v{S}palek}, \binits{R.}}:
\bctitle{Quantum verification of matrix products}.
In: \bbtitle{Proceedings of the Seventeenth Annual ACM-SIAM Symposium on Discrete Algorithm}.
\bsertitle{SODA '06},
pp. \bfpage{880}--\blpage{889}.
\bpublisher{Society for Industrial and Applied Mathematics},
\blocation{USA}
(\byear{2006})
\end{bchapter}
\endbibitem

\bibitem{sahapra}
\begin{barticle}
\bauthor{\bsnm{Saha}, \binits{A.}},
\bauthor{\bsnm{Majumdar}, \binits{R.}},
\bauthor{\bsnm{Saha}, \binits{D.}},
\bauthor{\bsnm{Chakrabarti}, \binits{A.}},
\bauthor{\bsnm{Sur-Kolay}, \binits{S.}}:
\batitle{Asymptotically improved circuit for a $d$-ary grover's algorithm with advanced decomposition of the $n$-qudit toffoli gate}.
\bjtitle{Phys. Rev. A}
\bvolume{105},
\bfpage{062453}
(\byear{2022}).
\doiurl{10.1103/PhysRevA.105.062453}
\end{barticle}
\endbibitem

\bibitem{Muthukrishnan_2000}
\begin{botherref}
\oauthor{\bsnm{Muthukrishnan}, \binits{A.}},
\oauthor{\bsnm{Stroud}, \binits{C.R.}}:
Multivalued logic gates for quantum computation.
Physical Review A
\textbf{62}(5)
(2000).
\doiurl{10.1103/physreva.62.052309}
\end{botherref}
\endbibitem

\bibitem{9410395}
\begin{barticle}
\bauthor{\bsnm{Saha}, \binits{A.}},
\bauthor{\bsnm{Mandal}, \binits{S.B.}},
\bauthor{\bsnm{Saha}, \binits{D.}},
\bauthor{\bsnm{Chakrabarti}, \binits{A.}}:
\batitle{One-dimensional lazy quantum walk in ternary system}.
\bjtitle{IEEE Transactions on Quantum Engineering}
\bvolume{2},
\bfpage{1}--\blpage{12}
(\byear{2021}).
\doiurl{10.1109/TQE.2021.3074707}
\end{barticle}
\endbibitem

\bibitem{Wang_2020}
\begin{botherref}
\oauthor{\bsnm{Wang}, \binits{Y.}},
\oauthor{\bsnm{Hu}, \binits{Z.}},
\oauthor{\bsnm{Sanders}, \binits{B.C.}},
\oauthor{\bsnm{Kais}, \binits{S.}}:
Qudits and high-dimensional quantum computing.
Frontiers in Physics
\textbf{8}
(2020).
\doiurl{10.3389/fphy.2020.589504}
\end{botherref}
\endbibitem

\bibitem{qft}
\begin{barticle}
\bauthor{\bsnm{Cao}, \binits{Y.}},
\bauthor{\bsnm{Peng}, \binits{S.-G.}},
\bauthor{\bsnm{Zheng}, \binits{C.}},
\bauthor{\bsnm{Long}, \binits{G.}}:
\batitle{Quantum fourier transform and phase estimation in qudit system}.
\bjtitle{Communications in Theoretical Physics}
\bvolume{55},
\bfpage{790}--\blpage{794}
(\byear{2011}).
\doiurl{10.1088/0253-6102/55/5/11}
\end{barticle}
\endbibitem

\bibitem{Bocharov_2017}
\begin{botherref}
\oauthor{\bsnm{Bocharov}, \binits{A.}},
\oauthor{\bsnm{Roetteler}, \binits{M.}},
\oauthor{\bsnm{Svore}, \binits{K.M.}}:
Factoring with qutrits: Shor’s algorithm on ternary and metaplectic quantum architectures.
Physical Review A
\textbf{96}(1)
(2017).
\doiurl{10.1103/physreva.96.012306}
\end{botherref}
\endbibitem

\bibitem{Fan_2007}
\begin{bchapter}
\bauthor{\bsnm{Fan}, \binits{Y.}}:
\bctitle{A generalization of the deutsch-jozsa algorithm to multi-valued quantum logic}.
In: \bbtitle{37th International Symposium on Multiple-Valued Logic (ISMVL'07)},
p. \bfpage{12}.
\bpublisher{IEEE Computer Society},
\blocation{Los Alamitos, CA, USA}
(\byear{2007}).
\doiurl{10.1109/ISMVL.2007.3}.
\burl{https://doi.ieeecomputersociety.org/10.1109/ISMVL.2007.3}
\end{bchapter}
\endbibitem

\bibitem{Khan_2006}
\begin{barticle}
\bauthor{\bsnm{Khan}, \binits{F.S.}},
\bauthor{\bsnm{Perkowski}, \binits{M.}}:
\batitle{Synthesis of multi-qudit hybrid and d-valued quantum logic circuits by decomposition}.
\bjtitle{Theoretical Computer Science}
\bvolume{367}(\bissue{3}),
\bfpage{336}--\blpage{346}
(\byear{2006}).
\doiurl{10.1016/j.tcs.2006.09.006}
\end{barticle}
\endbibitem

\bibitem{Di_2013}
\begin{botherref}
\oauthor{\bsnm{Di}, \binits{Y.-M.}},
\oauthor{\bsnm{Wei}, \binits{H.R.}}:
Synthesis of multivalued quantum logic circuits by elementary gates.
Physical Review A
\textbf{87}(1)
(2013).
\doiurl{10.1103/physreva.87.012325}
\end{botherref}
\endbibitem

\bibitem{balu2017physical}
\begin{botherref}
\oauthor{\bsnm{Balu}, \binits{R.}},
\oauthor{\bsnm{Castillo}, \binits{D.}},
\oauthor{\bsnm{Siopsis}, \binits{G.}}:
Physical realization of topological quantum walks on IBM-Q and beyond
(2017)
\end{botherref}
\endbibitem

\bibitem{Yan753}
\begin{barticle}
\bauthor{\bsnm{Yan}, \binits{Z.}},
\bauthor{\bsnm{Zhang}, \binits{Y.-R.}},
\bauthor{\bsnm{Gong}, \binits{M.}},
\bauthor{\bsnm{Wu}, \binits{Y.}},
\bauthor{\bsnm{Zheng}, \binits{Y.}},
\bauthor{\bsnm{Li}, \binits{S.}},
\bauthor{\bsnm{Wang}, \binits{C.}},
\bauthor{\bsnm{Liang}, \binits{F.}},
\bauthor{\bsnm{Lin}, \binits{J.}},
\bauthor{\bsnm{Xu}, \binits{Y.}},
\bauthor{\bsnm{Guo}, \binits{C.}},
\bauthor{\bsnm{Sun}, \binits{L.}},
\bauthor{\bsnm{Peng}, \binits{C.-Z.}},
\bauthor{\bsnm{Xia}, \binits{K.}},
\bauthor{\bsnm{Deng}, \binits{H.}},
\bauthor{\bsnm{Rong}, \binits{H.}},
\bauthor{\bsnm{You}, \binits{J.Q.}},
\bauthor{\bsnm{Nori}, \binits{F.}},
\bauthor{\bsnm{Fan}, \binits{H.}},
\bauthor{\bsnm{Zhu}, \binits{X.}},
\bauthor{\bsnm{Pan}, \binits{J.-W.}}:
\batitle{Strongly correlated quantum walks with a 12-qubit superconducting processor}.
\bjtitle{Science}
\bvolume{364}(\bissue{6442}),
\bfpage{753}--\blpage{756}
(\byear{2019}).
\doiurl{10.1126/science.aaw1611}
\end{barticle}
\endbibitem

\bibitem{alderete2020quantum}
\begin{botherref}
\oauthor{\bsnm{Alderete}, \binits{C.H.}},
\oauthor{\bsnm{Singh}, \binits{S.}},
\oauthor{\bsnm{Nguyen}, \binits{N.H.}},
\oauthor{\bsnm{Zhu}, \binits{D.}},
\oauthor{\bsnm{Balu}, \binits{R.}},
\oauthor{\bsnm{Monroe}, \binits{C.}},
\oauthor{\bsnm{Chandrashekar}, \binits{C.M.}},
\oauthor{\bsnm{Linke}, \binits{N.M.}}:
Quantum walks and Dirac cellular automata on a programmable trapped-ion quantum computer
(2020)
\end{botherref}
\endbibitem

\bibitem{acasiete2020experimental}
\begin{botherref}
\oauthor{\bsnm{Acasiete}, \binits{F.}},
\oauthor{\bsnm{Agostini}, \binits{F.P.}},
\oauthor{\bsnm{Moqadam}, \binits{J.K.}},
\oauthor{\bsnm{Portugal}, \binits{R.}}:
Implementation of quantum walks on ibm quantum computers.
Quantum Information Processing
\textbf{19}(12)
(2020).
\doiurl{10.1007/s11128-020-02938-5}
\end{botherref}
\endbibitem

\bibitem{singh2020universal}
\begin{botherref}
\oauthor{\bsnm{Singh}, \binits{S.}},
\oauthor{\bsnm{Alderete}, \binits{C.H.}},
\oauthor{\bsnm{Balu}, \binits{R.}},
\oauthor{\bsnm{Monroe}, \binits{C.}},
\oauthor{\bsnm{Linke}, \binits{N.M.}},
\oauthor{\bsnm{Chandrashekar}, \binits{C.M.}}:
Universal one-dimensional discrete-time quantum walks and their implementation on near term quantum hardware
(2020)
\end{botherref}
\endbibitem

\bibitem{Zhou_2019}
\begin{botherref}
\oauthor{\bsnm{Zhou}, \binits{J.-Q.}},
\oauthor{\bsnm{Cai}, \binits{L.}},
\oauthor{\bsnm{Su}, \binits{Q.-P.}},
\oauthor{\bsnm{Yang}, \binits{C.-P.}}:
Protocol of a quantum walk in circuit qed.
Physical Review A
\textbf{100}(1)
(2019).
\doiurl{10.1103/physreva.100.012343}
\end{botherref}
\endbibitem

\bibitem{santha}
\begin{botherref}
\oauthor{\bsnm{Santha}, \binits{M.}}:
Quantum walk based search algorithms
(2008).
\doiurl{10.48550/ARXIV.0808.0059}
\end{botherref}
\endbibitem

\bibitem{ambainis05}
\begin{botherref}
\oauthor{\bsnm{Ambainis}, \binits{A.}}:
Quantum search algorithms
(2005).
\doiurl{10.48550/ARXIV.QUANT-PH/0504012}
\end{botherref}
\endbibitem

\bibitem{ambainis2011search}
\begin{bchapter}
\bauthor{\bsnm{Ambainis}, \binits{A.}},
\bauthor{\bsnm{Backurs}, \binits{A.}},
\bauthor{\bsnm{Nahimovs}, \binits{N.}},
\bauthor{\bsnm{Ozols}, \binits{R.}},
\bauthor{\bsnm{Rivosh}, \binits{A.}}:
\bctitle{Search by quantum walks on two-dimensional grid without amplitude amplification},
vol. \bseriesno{7582}
(\byear{2011}).
\doiurl{10.1007/978-3-642-35656-8\_7}
\end{bchapter}
\endbibitem

\bibitem{Rhodes_2020}
\begin{botherref}
\oauthor{\bsnm{Rhodes}, \binits{M.L.}},
\oauthor{\bsnm{Wong}, \binits{T.G.}}:
Search on vertex-transitive graphs by lackadaisical quantum walk.
Quantum Information Processing
\textbf{19}(9)
(2020).
\doiurl{10.1007/s11128-020-02841-z}
\end{botherref}
\endbibitem

\bibitem{barenco}
\begin{barticle}
\bauthor{\bsnm{Barenco}, \binits{A.}},
\bauthor{\bsnm{Bennett}, \binits{C.H.}},
\bauthor{\bsnm{Cleve}, \binits{R.}},
\bauthor{\bsnm{DiVincenzo}, \binits{D.P.}},
\bauthor{\bsnm{Margolus}, \binits{N.}},
\bauthor{\bsnm{Shor}, \binits{P.}},
\bauthor{\bsnm{Sleator}, \binits{T.}},
\bauthor{\bsnm{Smolin}, \binits{J.A.}},
\bauthor{\bsnm{Weinfurter}, \binits{H.}}:
\batitle{Elementary gates for quantum computation}.
\bjtitle{Physical Review A}
\bvolume{52},
\bfpage{3457}--\blpage{3467}
(\byear{1995}).
\doiurl{10.1103/PhysRevA.52.3457}
\end{barticle}
\endbibitem

\bibitem{Garcia_Escartin_2013}
\begin{barticle}
\bauthor{\bsnm{Garcia-Escartin}, \binits{J.C.}},
\bauthor{\bsnm{Chamorro-Posada}, \binits{P.}}:
\batitle{A swap gate for qudits}.
\bjtitle{Quantum Information Processing}
\bvolume{12}(\bissue{12}),
\bfpage{3625}--\blpage{3631}
(\byear{2013}).
\doiurl{10.1007/s11128-013-0621-x}
\end{barticle}
\endbibitem

\bibitem{ambainis2004coins}
\begin{bchapter}
\bauthor{\bsnm{Ambainis}, \binits{A.}},
\bauthor{\bsnm{Kempe}, \binits{J.}},
\bauthor{\bsnm{Rivosh}, \binits{A.}}:
\bctitle{Coins make quantum walks faster}.
In: \bbtitle{SODA '05}
(\byear{2005})
\end{bchapter}
\endbibitem

\bibitem{ambainis2001one}
\begin{barticle}
\bauthor{\bsnm{Douglas}, \binits{B.L.}},
\bauthor{\bsnm{Wang}, \binits{J.B.}}:
\batitle{Efficient quantum circuit implementation of quantum walks}.
\bjtitle{Phys. Rev. A}
\bvolume{79},
\bfpage{052335}
(\byear{2009}).
\doiurl{10.1103/PhysRevA.79.052335}
\end{barticle}
\endbibitem

\bibitem{saha2020asymptotically}
\begin{botherref}
\oauthor{\bsnm{Saha}, \binits{A.}},
\oauthor{\bsnm{Majumdar}, \binits{R.}},
\oauthor{\bsnm{Saha}, \binits{D.}},
\oauthor{\bsnm{Chakrabarti}, \binits{A.}},
\oauthor{\bsnm{Sur-Kolay}, \binits{S.}}:
Asymptotically Improved Grover's Algorithm in any Dimensional Quantum System with Novel Decomposed $n$-qudit Toffoli Gate
(2020)
\end{botherref}
\endbibitem

\bibitem{gokhale2019asymptotic}
\begin{bchapter}
\bauthor{\bsnm{Gokhale}, \binits{P.}},
\bauthor{\bsnm{Baker}, \binits{J.M.}},
\bauthor{\bsnm{Duckering}, \binits{C.}},
\bauthor{\bsnm{Brown}, \binits{N.C.}},
\bauthor{\bsnm{Brown}, \binits{K.R.}},
\bauthor{\bsnm{Chong}, \binits{F.T.}}:
\bctitle{Asymptotic improvements to quantum circuits via qutrits}.
In: \bbtitle{Proceedings of the 46th International Symposium on Computer Architecture},
pp. \bfpage{554}--\blpage{566}
(\byear{2019})
\end{bchapter}
\endbibitem

\bibitem{IBMnoise}
\begin{barticle}
\bauthor{\bsnm{Fischer}, \binits{L.E.}},
\bauthor{\bsnm{Miller}, \binits{D.}},
\bauthor{\bsnm{Tacchino}, \binits{F.}},
\bauthor{\bsnm{Barkoutsos}, \binits{P.K.}},
\bauthor{\bsnm{Egger}, \binits{D.J.}},
\bauthor{\bsnm{Tavernelli}, \binits{I.}}:
\batitle{Ancilla-free implementation of generalized measurements for qubits embedded in a qudit space}.
\bjtitle{Phys. Rev. Res.}
\bvolume{4},
\bfpage{033027}
(\byear{2022}).
\doiurl{10.1103/PhysRevResearch.4.033027}
\end{barticle}
\endbibitem

\bibitem{Srikara_2020}
\begin{botherref}
\oauthor{\bsnm{Srikara}, \binits{S.}},
\oauthor{\bsnm{Chandrashekar}, \binits{C.M.}}:
Quantum direct communication protocols using discrete-time quantum walk.
Quantum Information Processing
\textbf{19}(9)
(2020).
\doiurl{10.1007/s11128-020-02793-4}
\end{botherref}
\endbibitem

\bibitem{Saha_2022}
\begin{botherref}
\oauthor{\bsnm{Saha}, \binits{A.}},
\oauthor{\bsnm{Majumdar}, \binits{R.}},
\oauthor{\bsnm{Saha}, \binits{D.}},
\oauthor{\bsnm{Chakrabarti}, \binits{A.}},
\oauthor{\bsnm{Sur-Kolay}, \binits{S.}}:
Faster search of clustered marked states with lackadaisical quantum walks.
Quantum Information Processing
\textbf{21}(8)
(2022).
\doiurl{10.1007/s11128-022-03606-6}
\end{botherref}
\endbibitem

\bibitem{8342181}
\begin{bchapter}
\bauthor{\bsnm{Zulehner}, \binits{A.}},
\bauthor{\bsnm{Paler}, \binits{A.}},
\bauthor{\bsnm{Wille}, \binits{R.}}:
\bctitle{Efficient mapping of quantum circuits to the ibm qx architectures}.
In: \bbtitle{2018 Design, Automation Test in Europe Conference Exhibition (DATE)},
pp. \bfpage{1135}--\blpage{1138}
(\byear{2018}).
\doiurl{10.23919/DATE.2018.8342181}
\end{bchapter}
\endbibitem

\end{thebibliography}

\begin{appendices}

\section{ {Quantum circuit for implementing the one-dimensional quantum walks in binary systems}}\label{appen1}

\begin{figure}[ht!]
\centering
\includegraphics[width=7cm, scale=1]{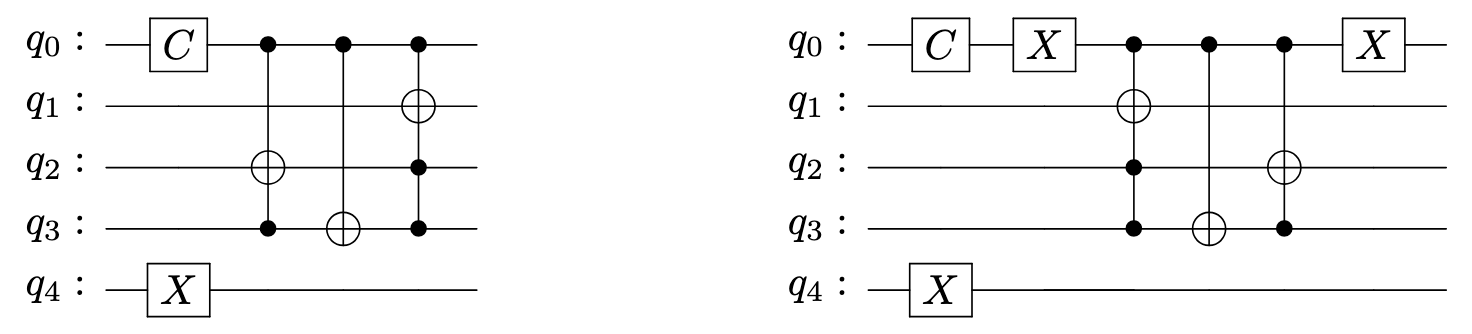}
\captionof{figure}{{Quantum circuit on (left) and (right) for implementing at most seven steps of DTQW on five qubits as given in Figure \ref{tabbinarywalk}. If the initial state is even, the circuit on (left) is applied first, and if the initial state is odd, the circuit on (right) is applied first, where even and odd are determined by the value of the last qubit.}}
\label{binarywalk}
\end{figure}

\begin{figure}[ht!]
\centering
\includegraphics[width=8cm, scale=1]{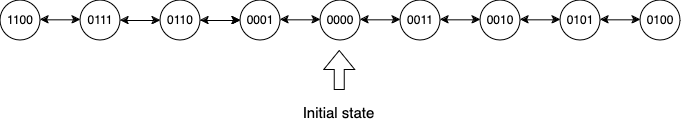}
\captionof{figure}{{Position state mapping with the multi-qubits states for quantum circuits presented in Figure \ref{binarywalk}.}}
\label{tabbinarywalk}
\end{figure}

In \cite{ambainis2001one}, the authors first showed the well-known increment-decrement approach for DTQW implementation. In \cite{singh2020universal}, the authors presented the logical realization of quantum circuits to implement one-dimensional quantum walks using the two-dimensional coin 4-qubit quantum systems with the help of the nearest-neighbor approach of position mapping. Albeit, unfortunately, there is no generic nearest-neighbor approach for DTQW implementation for $n$-qubit systems. To implement a DTQW in a one-dimensional position Hilbert space of size $2^q$, $(q)$ qubits and one qubit are required, one qubit to represent the particle's internal state (coin qubit) and $q$-qubits to represent its position. The coin operation can be implemented by applying a single qubit rotation gate on the coin qubit, and the position shift operation is implemented subsequently with the help of multi-qubit gates where the coin qubit acts as the control. Quantum circuits for implementing DTQW depend on how the position space is represented. As an example, Figure \ref{binarywalk} portrays the quantum circuits for implementing the first seven steps of one-dimensional DTQW on five qubits. These circuits are based on the appropriate position state mapping with multi-qubits states for one-dimensional DTQW, which is shown in Figure \ref{tabbinarywalk}. These circuits are needed to implement alternatively by considering if the
initial state is even, circuit \ref{binarywalk} on the left is applied first, and if the initial state
is odd circuit \ref{binarywalk} on the right is applied first, where even and odd determined
by the value of the last qubit. The authors further exhibited that with the help of a higher-controlled multi-controlled Toffoli gate, these circuits can be scaled to more steps of DTQW on line. 

\section{ {Quantum circuit for implementing the one-dimensional quantum walks in ternary systems} }\label{appen2}
 In \cite{8342181}, the authors showed the logical realization of quantum circuits to implement one-dimensional quantum walks using two-dimensional coins in ternary quantum systems. Similar to binary systems, to implement a DTQW in a one-dimensional position Hilbert space of size $3^q$, $(q)$ qutrits and one qubit are required, one qubit to represent the particle's internal state (coin qubit) and $q$-qutrits to represent its position. Example circuits for four qubit-qutrit systems are given in this section.

\begin{figure}[ht!]
\centering
\includegraphics[width=8cm, scale=1]{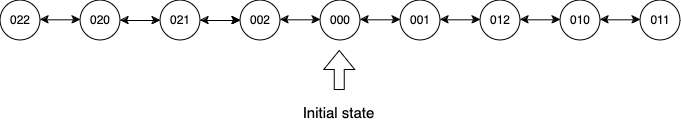}
\captionof{figure}{{Position state mapping with the multi-qutrits states for quantum circuits presented in Figures \ref{fig6} and \ref{fig7}.}}
\label{tab3}
\end{figure}

For $q=3$, the number of steps of DTQW that can be implemented is $\lfloor3^{q}/2\rfloor = 13$. The authors of \cite{9410395} have chosen the position state mapping given in Figure \ref{tab3} with a fixed initial position state $\ket{000}$. Fixing the initial state of the walker helps in reducing the gate count in the quantum circuit and hence reduces the overall error. After each step of the DTQW, two new position states must be considered. In Figure \ref{tab3}, the authors showed that the mapping of these new position states onto the multi-qutrit states is in such a way that the efficient number of gates are used to implement the shift operation, they consider the nearest-neighbor position space so as to make the circuit efficient.

\begin{figure}[ht!]
\centering
\includegraphics[width=4cm, scale=1]{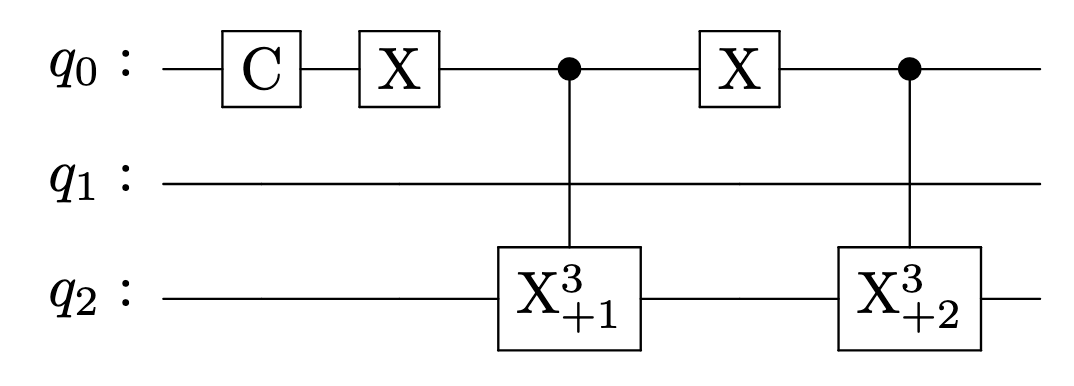}
\captionof{figure}{{Quantum circuit for the first step of DTQW on three qutrits as given in Figure \ref{tab3}.}}
\label{fig6}
\end{figure}

\begin{figure}[ht!]
\centering
\includegraphics[width=6cm, scale=1]{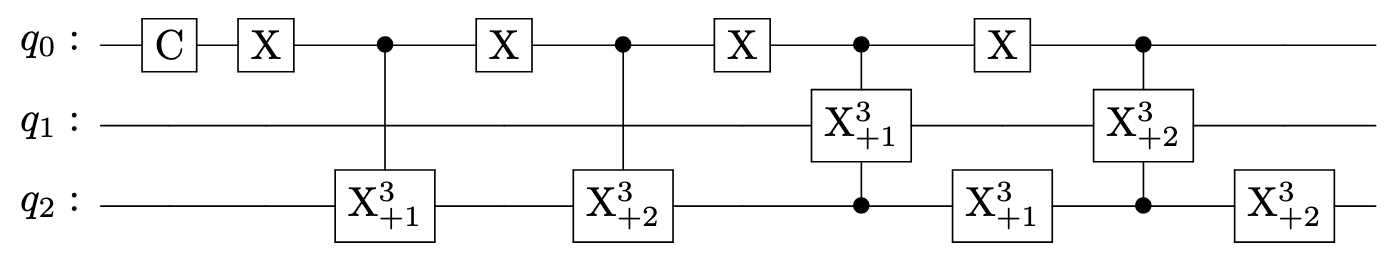}
\captionof{figure}{{Quantum circuit for the second-fourth steps of DTQW on three qutrits as shown in Figure \ref{tab3}.}}
\label{fig7}
\end{figure}

They have observed that in the first step of DTQW only one qutrit (last qutrit) change was enough to describe the position states as after the first step of DTQW, there are only three different position states in the one-dimensional line, which is portrayed in Figure \ref{fig6}. Similarly, for the next three steps, the change is required on two qutrits (second and last qutrit) to get all possible states to describe all four steps of DTQW. Similarly, for the fifth step of DTQW, two new position states are needed to be introduced. For that, the change in three qutrits is required. At this stage to move forward, the circuit design for previous position states remains the same as shown in Figure \ref{fig7}. For further next eight steps of DTQW, this same circuit works as no new qutrit is required to represent the position states since they are following the nearest-neighbor approach.

They also showed that these circuits can be scaled to implement more steps of one-dimensional DTQW on larger ternary systems with the help of higher controlled ternary gates. Using $n$-qutrit systems, implementation of $\lfloor3^{n}/2\rfloor$-steps of a DTQW can be performed. In other words, to implement the $n$-step of DTQW, at most $\lceil\log_{3}2n\rceil$ qutrits are required. 

\section{{Implementation of one-dimensional DTQW in 7-ary quantum systems}}\label{7-ary}
\begin{figure}[ht!]
\centering   
\includegraphics[scale=0.12]{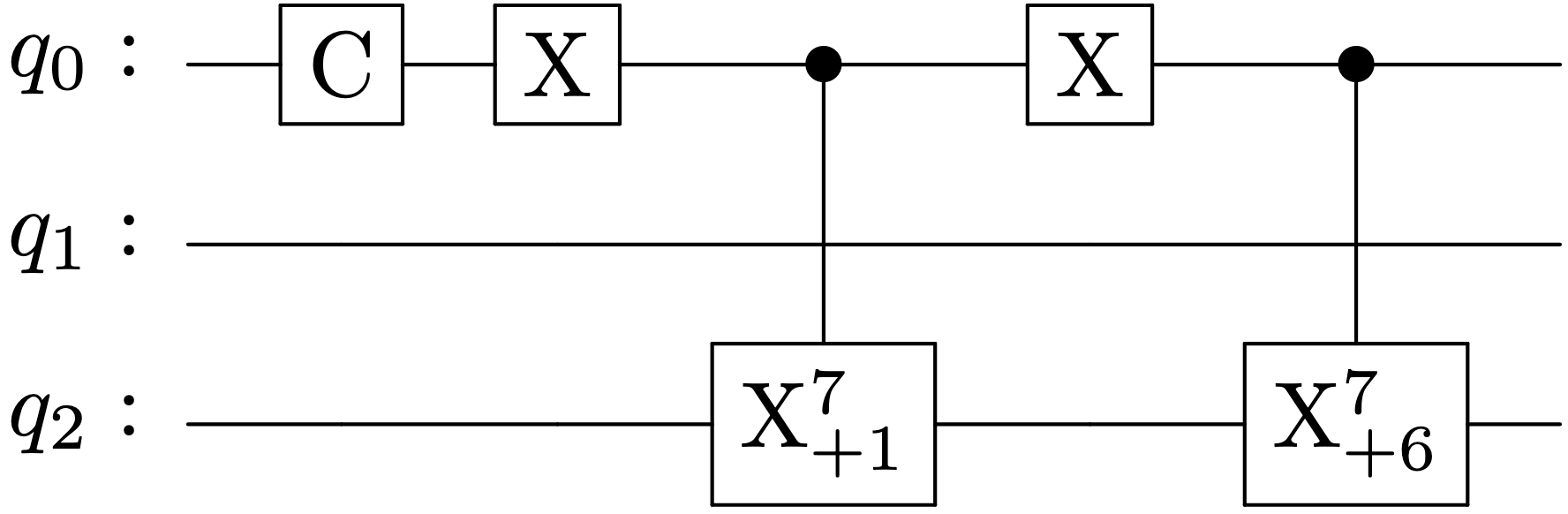}
    \caption{{Quantum circuit for the first three steps of DTQW on two qudits in 7-ary quantum systems as given in Table \ref{tab_7_dtqw}.}}
    \label{fig:71}
\end{figure}

\begin{figure}[ht!]
\centering   
\includegraphics[scale=0.15]{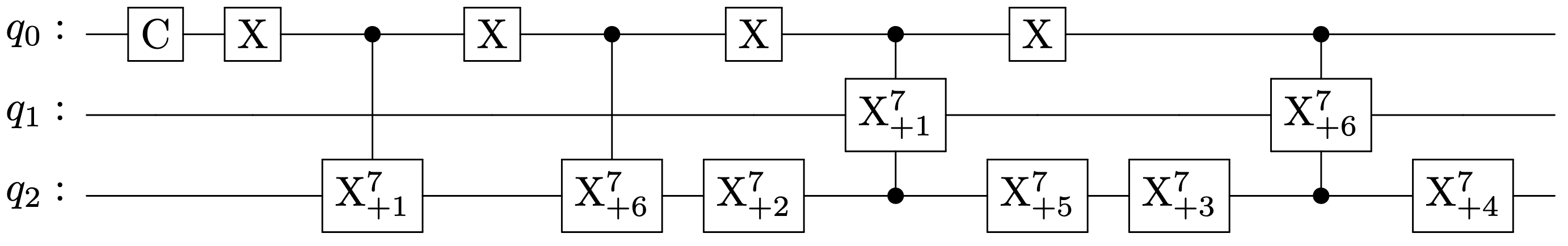}
    \caption{{Quantum circuit for the fourth-twenty fourth steps of DTQW on two qudits in 7-ary quantum systems as shown in Table \ref{tab_7_dtqw}.}}
    \label{fig:72}
\end{figure}

Here, we take a quantum system $i.e.,$ 7-ary quantum systems for the better understanding of the implementation of DTQW in odd dimensions. In Table \ref{tab_7_dtqw}, an example of position state mapping onto the multi-qudits states for the first twenty-four one-dimensional DTQW steps in 7-ary quantum systems has been presented. The corresponding quantum circuits for Table \ref{tab_7_dtqw} are portrayed in Figure \ref{fig:71} and Figure \ref{fig:72}. These two circuits presented in  Figure \ref{fig:71} and Figure \ref{fig:72} need to be executed to implement the first twenty-four DTQW steps. Figure \ref{fig:71} needs to be implemented three times for the first three steps of DTQW and for the rest of the DTQW steps as described in Table \ref{tab_7_dtqw}, we need to implement the quantum circuit as shown in Figure \ref{fig:72}. It can be inferred that to implement the first twenty-four steps of one-dimensional DTQW, two qudits are sufficient in 7-ary quantum systems as compared to the three qudits in 5-ary quantum systems.

\begin{table}[ht!]
\centering
\caption{An example of position state mapping onto the multi-qudits states in 7-ary quantum systems for quantum circuits presented in Figure \ref{fig:71}, Figure \ref{fig:72} and Figure \ref{fig:73}.}
\begin{tabular}{|m{11em} | m{0.005em}| m{11em}  | }
\hline
~~~$\ket{x = 0} \equiv \ket{00}$  &&   \\
\hline
~~~$\ket{x = 1} \equiv \ket{01}$ && ~~~$\ket{x = -1} \equiv \ket{06}$\\
\hline 
~~~$\ket{x = 2} \equiv \ket{02}$  &&  ~~~$\ket{x = -2} \equiv \ket{05}$ \\
\hline
~~~$\ket{x = 3} \equiv \ket{03}$ && ~~~$\ket{x = -3} \equiv \ket{04}$\\
\hline
~~~$\ket{x = 4} \equiv \ket{14}$  && ~~~$\ket{x = -4} \equiv \ket{63}$\\
\hline
~~~$\ket{x = 5} \equiv \ket{15}$ &&  ~~~$\ket{x = -5} \equiv \ket{62}$\\
\hline 
~~~$\ket{x = 6} \equiv \ket{16}$ && ~~~$\ket{x = -6} \equiv \ket{61}$ \\
\hline
~~~$\ket{x = 7} \equiv \ket{10}$ && ~~~$\ket{x = -7} \equiv \ket{60}$ \\
\hline
~~~$\ket{x = 8} \equiv \ket{11}$ && ~~~$\ket{x = -8} \equiv \ket{66}$ \\
\hline
~~~$\ket{x = 9} \equiv \ket{12}$ && ~~~$\ket{x = -9} \equiv \ket{65}$ \\
\hline
~~~$\ket{x = 10} \equiv \ket{13}$ && ~~~$\ket{x = -10} \equiv \ket{64}$ \\
\hline
~~~$\ket{x = 11} \equiv \ket{24}$ && ~~~$\ket{x = -11} \equiv \ket{53}$ \\
\hline
~~~$\ket{x = 12} \equiv \ket{25}$ && ~~~$\ket{x = -12} \equiv \ket{52}$ \\
\hline
~~~$\ket{x = 13} \equiv \ket{26}$ && ~~~$\ket{x = -13} \equiv \ket{51}$ \\

\hline
~~~$\ket{x = 14} \equiv \ket{20}$ && ~~~$\ket{x = -14} \equiv \ket{50}$\\
\hline 
~~~$\ket{x = 15} \equiv \ket{21}$  &&  ~~~$\ket{x = -15} \equiv \ket{56}$ \\
\hline
~~~$\ket{x = 16} \equiv \ket{22}$ && ~~~$\ket{x = -16} \equiv \ket{55}$\\
\hline
~~~$\ket{x = 17} \equiv \ket{23}$  && ~~~$\ket{x = -17} \equiv \ket{54}$\\
\hline
~~~$\ket{x = 18} \equiv \ket{34}$ &&  ~~~$\ket{x = -18} \equiv \ket{43}$\\
\hline 
~~~$\ket{x = 19} \equiv \ket{35}$ && ~~~$\ket{x = -19} \equiv \ket{42}$ \\
\hline
~~~$\ket{x = 20} \equiv \ket{36}$ && ~~~$\ket{x = -20} \equiv \ket{41}$ \\
\hline
~~~$\ket{x = 21} \equiv \ket{30}$ && ~~~$\ket{x = -21} \equiv \ket{40}$ \\
\hline
~~~$\ket{x = 22} \equiv \ket{31}$ && ~~~$\ket{x = -22} \equiv \ket{46}$ \\
\hline
~~~$\ket{x = 23} \equiv \ket{32}$ && ~~~$\ket{x = -23} \equiv \ket{45}$ \\
\hline
~~~$\ket{x = 24} \equiv \ket{33}$ && ~~~$\ket{x = -24} \equiv \ket{44}$ \\
\hline
\end{tabular}
\label{tab_7_dtqw}
\end{table}

\begin{figure}[ht!]
\centering   
\includegraphics[scale=0.15]{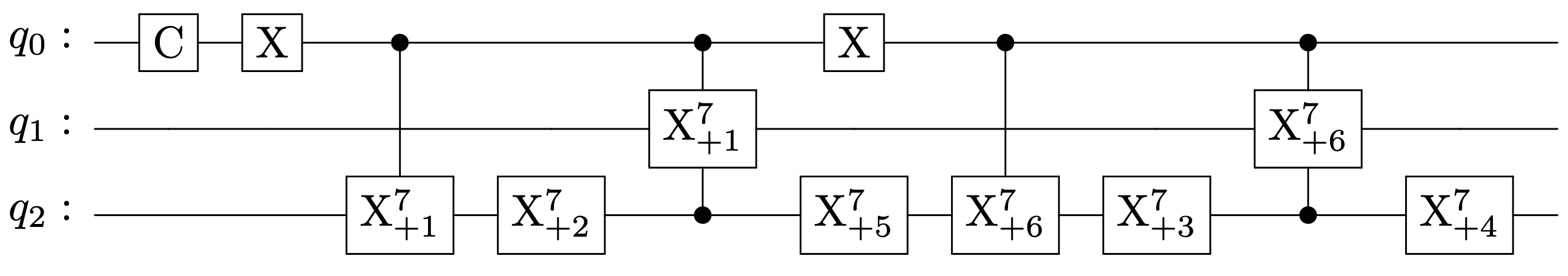}
    \caption{{Optimized quantum circuit for the fourth-twenty fourth steps of DTQW on two qudits in 7-ary quantum systems as shown in Table \ref{tab_7_dtqw}.}}
    \label{fig:73}
\end{figure}

In Figure \ref{fig:72}, a further optimization on the quantum circuit has been carried out, which is portrayed in Figure \ref{fig:73}. After applying a conventional one-qubit/qudit gate reduction identity rule \cite{barenco}, in Figure \ref{fig:73}, we can eliminate one one-qubit/qudit gate. These one-qubit/qudit gate reduction rules can be further applied to the proposed circuit for optimization.

\section{{Implementation of one-dimensional DTQW in 6-ary quantum systems}}\label{6-ary}
\begin{figure}[ht!]
  \centering 
\includegraphics[scale=0.2]{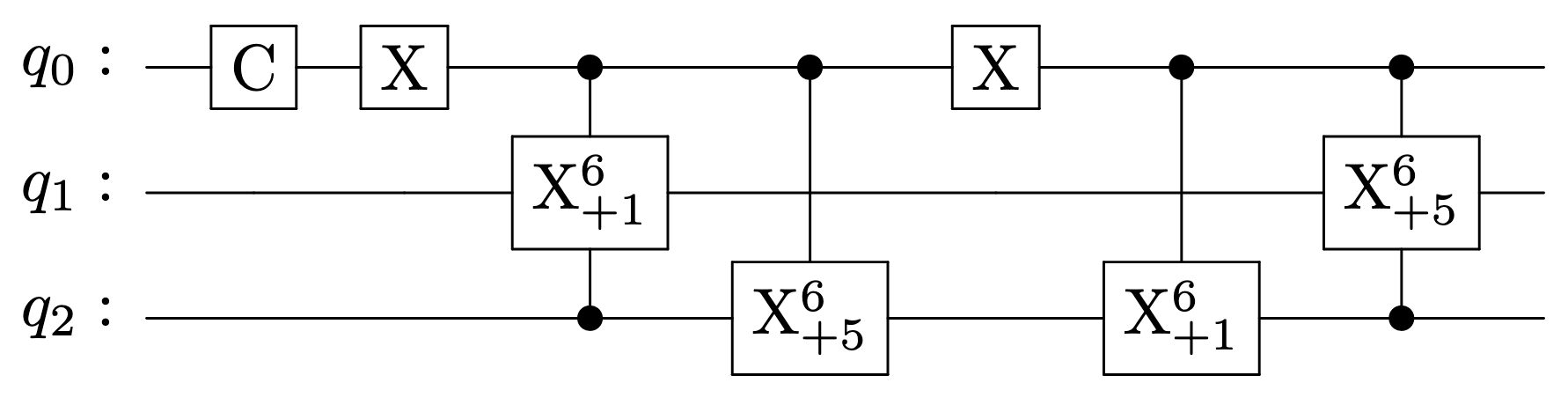}
    \caption{{Quantum circuit for all the seventeen steps of DTQW on two qudits in 6-ary quantum systems as shown in Table \ref{tab_6_dtqw}.}}
    \label{fig:61}
\end{figure}

Here, we consider a quantum system $i.e.,$ 6-ary quantum systems for the better understanding of the implementation of DTQW in even dimensions. In Table \ref{tab_6_dtqw}, an example of position state mapping onto the multi-qudits states for the first seventeen one-dimensional DTQW steps in 6-ary quantum systems has been presented. It can also be inferred that to implement seventeen steps of one-dimensional DTQW, two qudits are sufficient in 6-ary quantum systems. Two new position states must be taken into consideration in this situation as well after each DTQW step. The corresponding quantum circuits for Table \ref{tab_6_dtqw} are portrayed in Figure \ref{fig:61}. 

    \begin{table}[ht!]
\centering
\caption{An example of position state mapping onto the multi-qudits states in 6-ary quantum systems.}
\begin{tabular}{|m{11em} | m{0.005em}| m{11em}  | }
\hline
~~~$\ket{x = 0} \equiv \ket{00}$  &&   \\
\hline
~~~$\ket{x = 1} \equiv \ket{01}$ && ~~~$\ket{x = -1} \equiv \ket{55}$\\
\hline 
~~~$\ket{x = 2} \equiv \ket{02}$  &&  ~~~$\ket{x = -2} \equiv \ket{54}$ \\
\hline
~~~$\ket{x = 3} \equiv \ket{03}$ && ~~~$\ket{x = -3} \equiv \ket{53}$\\
\hline
~~~$\ket{x = 4} \equiv \ket{04}$  && ~~~$\ket{x = -4} \equiv \ket{52}$\\
\hline
~~~$\ket{x = 5} \equiv \ket{05}$ &&  ~~~$\ket{x = -5} \equiv \ket{51}$\\
\hline 
~~~$\ket{x = 6} \equiv \ket{10}$ && ~~~$\ket{x = -6} \equiv \ket{50}$ \\
\hline
~~~$\ket{x = 7} \equiv \ket{11}$ && ~~~$\ket{x = -7} \equiv \ket{45}$ \\
\hline
~~~$\ket{x = 8} \equiv \ket{12}$ && ~~~$\ket{x = -8} \equiv \ket{44}$ \\
\hline
~~~$\ket{x = 9} \equiv \ket{13}$ && ~~~$\ket{x = -9} \equiv \ket{43}$ \\
\hline
~~~$\ket{x = 10} \equiv \ket{14}$ && ~~~$\ket{x = -10} \equiv \ket{42}$ \\
\hline
~~~$\ket{x = 11} \equiv \ket{15}$ && ~~~$\ket{x = -11} \equiv \ket{41}$ \\
\hline
~~~$\ket{x = 12} \equiv \ket{20}$ && ~~~$\ket{x = -12} \equiv \ket{40}$ \\
\hline
~~~$\ket{x = 13} \equiv \ket{21}$ && ~~~$\ket{x = -13} \equiv \ket{35}$ \\
\hline
~~~$\ket{x = 14} \equiv \ket{22}$ && ~~~$\ket{x = -14} \equiv \ket{34}$\\
\hline 
~~~$\ket{x = 15} \equiv \ket{23}$  &&  ~~~$\ket{x = -15} \equiv \ket{33}$ \\
\hline
~~~$\ket{x = 16} \equiv \ket{24}$ && ~~~$\ket{x = -16} \equiv \ket{32}$\\
\hline
~~~$\ket{x = 17} \equiv \ket{25}$  && ~~~$\ket{x = -17} \equiv \ket{31}$\\
\hline
\end{tabular}
\label{tab_6_dtqw}
\end{table}

\section{{Implementation of two-dimensional DTQW in 5-ary quantum systems}}\label{2d_appendix}

\begin{figure}[ht!]
\centering
\includegraphics[width=4cm, scale=1]{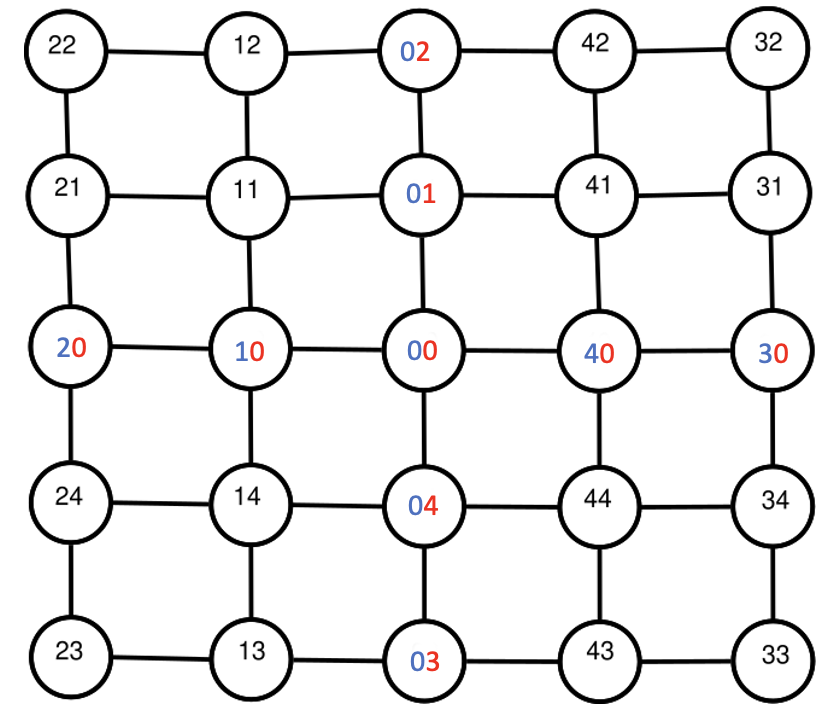}
\captionof{figure}{{An example of position state mapping onto the multi-qudits states in 5-ary quantum systems for quantum circuits presented in Figure \ref{fig:5_2d}. The first qudit (color in blue) is considered for horizontal one-dimensional quantum walks, and the last qudit (color in red) is considered for vertical one-dimensional quantum walks.}}
\label{2d_5}
\end{figure}

\begin{figure}[ht!]
\centering   
\includegraphics[scale=0.5]{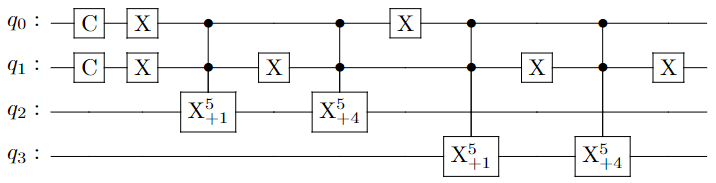}
    \caption{Quantum circuit for the first two steps of two-dimensional DTQW on two qudits in 5-ary quantum systems as shown in Figure \ref{2d_5}.}
    \label{fig:5_2d}
\end{figure}

In Figure \ref{2d_5}, an example of position state mapping onto the multi-qudits states for the first two two-dimensional DTQW steps in 5-ary quantum systems has been presented. The corresponding quantum circuit for Figure \ref{2d_5} is portrayed in Figure \ref{fig:5_2d}. The circuit presented in  Figure \ref{fig:5_2d} has to be executed two times consecutively to implement the first two two-dimensional DTQW steps. It can also be inferred that to implement the first two steps of two-dimensional DTQW, two qudits are sufficient in 5-ary quantum systems as compared to the four qudits in 3-ary quantum systems. 

\end{appendices}

\end{document}